\documentclass[aps, nofootinbib,superscriptaddress]{revtex4}
\usepackage{psfrag}
\usepackage{array}
\usepackage{amstext,amsmath,amssymb,amsfonts,bbm}
\usepackage[latin1]{inputenc}
\usepackage[dvips]{graphicx}
\usepackage{epsfig}

\newcommand{\f}{\frac}
\newcommand{\tr}{\mathrm{tr}}

\newcommand{\su}{\mathfrak{su}}

\renewcommand{\u}{{\mathfrak u}}

\newcommand{\spin}{\mathfrak{spin}}

\def\la{\langle}
\def\ra{\rangle}

\newcommand{\N}{\mathbb{N}}
\newcommand{\Z}{\mathbb{Z}}

\newcommand{\R}{\mathbb{R}}

\newcommand{\C}{\mathbb{C}}

\newcommand{\cF}{{\mathcal F}}

\newcommand{\cH}{{\mathcal H}}

\newcommand{\cO}{{\mathcal O}}
\newcommand{\cM}{{\mathcal M}}
\newcommand{\cN}{{\mathcal N}}

\newcommand{\cC}{{\mathcal C}}
\newcommand{\cS}{{\mathcal S}}
\newcommand{\SU}{\mathrm{SU}}
\newcommand{\SL}{\mathrm{SL}}
\newcommand{\GL}{\mathrm{GL}}
\newcommand{\SO}{\mathrm{SO}}
\newcommand{\U}{\mathrm{U}}

\newcommand{\Spin}{\mathrm{Spin}}

\newcommand{\vJ}{\vec{J}}
\newcommand{\vV}{\vec{V}}
\def\bv{{\bf v}}
\newcommand{\w}{\wedge}
\def\tDelta{{\widetilde{\Delta}}}

\newcommand{\id}{\mathbb{I}}

\def\eps{\epsilon}
\def\tl{\widetilde}

\def\tz{\tl{z}}
\def\tX{\tl{X}}
\def\arr{\rightarrow}

\def\nn{\nonumber}

\newcommand{\be}{\begin{equation}}
\newcommand{\ee}{\end{equation}}
\newcommand{\bes}{\begin{eqnarray}}
\newcommand{\ees}{\end{eqnarray}}
\newcommand{\bea}{\begin{eqnarray}}
\newcommand{\eea}{\end{eqnarray}}

\newcommand{\Ref}[1]{(\ref{#1})}

\newcommand{\mat}[2]{\left(\begin{array}{#1}#2
\end{array}\right)}

\def\tj{{\widetilde{j}}}

\def\pp{\partial}

\def\dag{^\dagger}

\newcommand{\bin} [2] {\left (\begin{array}{c}#2\\#1\end{array} \right ) }
\begin{document}

\title{\large\bf Revisiting the Simplicity Constraints and Coherent Intertwiners}

\author{{\bf Ma\"\i t\'e Dupuis}\email{maite.dupuis@ens-lyon.fr}}
\affiliation{Laboratoire de Physique, ENS Lyon, CNRS-UMR 5672, 46 All\'ee d'Italie, Lyon 69007, France}

\author{{\bf Etera R. Livine}\email{etera.livine@ens-lyon.fr}}
\affiliation{Laboratoire de Physique, ENS Lyon, CNRS-UMR 5672, 46 All\'ee d'Italie, Lyon 69007, France}

\date{\small \today}

\begin{abstract}

%In this work, we study the $\U(N)$ coherent states and the action of the different SU(2) invariant operators on them.
%We then use the $\U(N)$ tools to revisit how to impose the simplicity constraints on intertwiners for 4d Euclidean gravity in the spinfoam framework. We show that the discrete cross simplicity constraints can be replaced by strong constraints. Solving these new constraints allow to build new intertwiners from the $\U(N)$ coherent states which then solve weakly all simplicity constraints for an arbitrary Immirzi parameter.

In the context of loop quantum gravity and spinfoam models, the simplicity constraints are essential in that they allow to write general relativity as a constrained topological BF theory. In this work, we apply the recently developed $\U(N)$ framework for $\SU(2)$ intertwiners to the issue of imposing the simplicity constraints to spin network states. More particularly, we focus on solving them on individual intertwiners in the 4d Euclidean theory. We review the standard way of solving the simplicity constraints using coherent intertwiners and we explain how these fit within the $\U(N)$ framework. Then we show how these constraints can be written as a closed $\u(N)$ algebra and we propose a set of $\U(N)$ coherent states that solves all the simplicity constraints weakly for an arbitrary Immirzi parameter.

\end{abstract}

\maketitle
%%%%%%%%%%%%%%%%%%%%%%%%%%%%%%%%%%%%%%%%%

%%%%%%
\section*{Introduction}
%%%%%%
%
%\begin{itemize}
%
%\item Spinfoam from discretized the path integral for topological BF theory plus constraints.
%
%\item Simplicity constraints to restrict the $B$-field to come from a tetrad field, thus ``reducing" BF theory to general relativity
%
%\item Simplicity constraints are 2nd class. How to impose them on the path integral? At the discrete level?
%
%\item First try, the Barrett-Crane model to impose them strongly, drawback: no intertwiner dynamics
%
%\item Then relaxing the constraints and imposing them weakly using coherent intertwiner states?
%
%\item Implemented explicitly similarly to the Gupta-Bleuer quantization in the EPRL-FK spinfoam models
%
%\item Here, using the newly developed $\U(N)$ framework, closed alg of invariant operators acting on the intertwiner space. Could turn out useful?
%
%\item Give plan of paper
%
%\end{itemize}

Spinfoam models present a tentative framework for a regularized path integral for quantum gravity. Initially constructed as a history formalism for loop quantum gravity describing the evolution of spin network states of geometry, it has been developed since then as a discretization of the path integral of general relativity reformulated as a topological BF gauge theory plus constraints. Since the BF theory is topological, it does not have any local degree of freedom and can be discretized and quantized exactly as a spinfoam model without losing any of its physical content. Then one works directly at the quantum level in the spinfoam framework and attempts to impose consistently the constraints turning the BF theory into a geometrical theory and introducing local degrees of freedom.

More precisely, the Plebanski action writes general relativity as a constrained BF gauge theory for the Lorentz group $\SO(3,1)$ (or $\SO(4)$ in the Euclidean case):
%\be
$$
S_{\textrm{GR}}[B, \omega, \lambda]= \int_\cM B^{IJ} \wedge F_{IJ}[\omega] + \lambda_\alpha \mathcal{C}_\alpha [B],
$$
%\ee
where $\cM$ is the space-time manifold, $I,J$ are Lorentz indices  running from 0 to 3, $\omega$ is a  $\mathfrak{so}(3,1)$-valued 1-form (or $\mathfrak{so}(4)$-valued in the Euclidean theory) and $F[\omega]$ is its strength tensor, $B$ is a $\mathfrak{so}(3,1)$ valued 2-form.
The constraints $\mathcal{C}_\alpha[B]$ are enforced by the Lagrange multipliers $\lambda_\alpha$ and are called ``simplicity" constraints. They are usually quadratic in the $B$-field and constrain it to come from a tetrad field $e$ in a such way that we recover general relativity in its first order formalism formulated in term of tetrad and Lorentz connection (see e.g. \cite{plebanski}). This reformulation is at the heart of the spinfoam models developed up to now. Another approach is based on the MacDowell-Mansouri action, which writes general relativity as a BF theory for the gauge group $\SO(4,1)$ (or $\SO(5)$ in the Euclidean case) with a non-trivial potential in the $B$-field which breaks the symmetry down back to the Lorentz group \cite{artem}. Although this is a very interesting alternative, it hasn't yet lead to a definite proposal for a spinfoam model. We will therefore focus on spinfoam models attempting to implement a discrete path integral for the constrained BF theory with gauge group $\SO(3,1)$ or $\SO(4)$.

At the continuum level, the simplicity constraints  are essential in turning the non-physical BF theory into 4d gravity. They are second class constraints and modify non-trivially the path integral \cite{continuum1,continuum2,continuum3}. At the spinfoam level, we usually start with a triangulation of the 4d space-time manifold (or more generally, any cellular decomposition of the manifold) and we discretize the fields, both the $B$-field and the Lorentz gauge connection. The $B$-field is a 2-form and is naturally discretized on the faces of the triangulation, i.e the triangles. In the spinfoam framework, the discrete $B$-field is then quantized into a representation of the gauge group associated to each triangle. Next, the simplicity constraints can be discretized as well and are usually solved separately on each 3-cell of the triangulation i.e the tetrahedra. Upon the spinfoam quantization procedure, a tetrahedron becomes an intertwiner (tensor invariant under the gauge group) between the four representations living on its four triangles. Finally, the simplicity constraints are to be imposed on these intertwiners. At this discretized and quantized level, a big issue is that the algebra of these simplicity constraints does not close. This reflects the fact that they correspond to second class constraints.

The first explicit attempt to solve these discrete simplicity constraints lead to the Barrett-Crane spinfoam model, which was actually initially built as a geometrical quantization of individual 4-simplices \cite{BC,BC2}. In this case, the constraints were imposed strongly and a single solution was found \cite{unique}: to each tetrahedron is associated the unique Barrett-Crane intertwiner (once the four representations living on the associated triangles are fixed). Although it seems to be a happy co\"\i ncidence to find such a unique solution, it also has the drawback that it seems to freeze too many degrees of freedom of the 3d space geometry, especially considered from the point of view of loop quantum gravity or the spinfoam graviton calculations \cite{alesci}.

It was then proposed that the simplicity should be imposed weakly, using coherent states \cite{coh1} or following a procedure similar to the Gupta-Bleuler quantization \cite{EPR0}. These two possibilities were shown to lead to the same family of models, the EPRL-FK spinfoam models \cite{EPR,FK,LS,EPRL}, which rely heavily on the use of coherent intertwiners \cite{coh1,coh2,coh3}.

In the present paper, we would like to revisit the implementation of the discrete simplicity constraints from the perspective of the recently developed $\U(N)$ framework for ($\SU(2)$) intertwiners \cite{UN1,UN2,UN3}. This framework proposes a closed algebra of geometric observables acting on the space of intertwiners. It is thus natural to wonder if the simplicity constraints can be recast in term of these observables and see if it allows to propose a closed algebra of simplicity constraints.

We will restrict ourselves to the study of the Euclidean case, i.e constrained BF theory with gauge group $\Spin(4)$ (which is the double cover of $\SO(4)$) and the corresponding intertwiners. We have not yet investigated how our approach can be applied to the Lorentzian case yet. We will focus primarily on the theory without Immirzi parameter, but our final proposal will hold for an arbitrary value of the Immirzi parameter. We will start by reviewing the discrete simplicity constraints and how to solve them in term of coherent intertwiners. Then we will introduce the $\U(N)$ framework and explain how to reformulate the coherent intertwiners in this context. We analyze in details the operators acting on these coherent intertwiners. In the third part, we will apply  the $\U(N)$ tools to the issue of the simplicity constraints: we propose  new sets of simplicity constraints which form  closed algebra and therefore can be solved strongly. However, this will hint towards the fact that enforcing strongly the diagonal simplicity constraints is too restrictive and physically unjustified. In the last section, we will therefore relax the diagonal simplicity constraints and this will allow us to solve weakly all the simplicity constraints for an arbitrary Immirzi parameter in term of the $\U(N)$ coherent states.

%%%%%%%%%
\section{Introducing the Simplicity Constraints}
%%%%%%%%%
%%%%%%

%%%
\subsection{Discretizing Polyhedra into Intertwiners}
%%%

The spinfoam framework is based on discretized space-time manifolds, built from gluing 4-cells together. The key kinematical ingredient is the boundary of the 4-cells. More precisely, the boundary of a given 4-cell is made of 3-cells (usually tetrahedra) glued together. To each 3-cell is associated an intertwiner for the Lorentz group (or more generally for the considered gauge group). This intertwiner lives on the dual 1-skeleton of the 3-cell: the intertwiner vertex corresponds to the inside of the 3-cell while each of the  intertwiners legs pierces one (2d) face of the 3-cell. Then the intertwiners associated to all these 3-cells are glued together into one spin network which defines the quantum state of the boundary of the 4-cell. Finally the standard prescription for the spinfoam amplitude  associated to a given 4-cell is to evaluate its boundary spin network on the flat connection (all holonomies set to the identity). Then the spinfoam amplitude for the whole discretized space-time is defined as a sum over intertwiners with a specific statistical weight of the product of the spinfoam 4-cell amplitudes. To sum up this construction, the key step in the spinfoam quantization is to define an intertwiner corresponding to each 3-cell. This is where the discrete simplicity constraints usually come into play.

Let us therefore focus on one 3-cell. In the following, we will work in the Euclidean theory, with the $\spin(4)$ Lie algebra. The $B$-field is discretized on the faces of this (3d) polyhedron by associating a bivector $B^{IJ}_\Delta$ to each face $\Delta$ on the polyhedron boundary. $I,J$ are Lorentz indices and run from 0 to 3.  These bivectors define the geometry of the polyhedron. We distinguish two sectors: the normal bivector to the face (the normal to a plane embedded in a 4d manifold is indeed a bivector) is either $B^{IJ}_\Delta$, which we call the standard sector $(s)$, or it is given by its Hodge dual $(\star B)^{IJ}_\Delta\,\equiv\f12 \epsilon^{IJ}{}_{KL}B^{KL}_\Delta$, which we call the dual sector $(\star)$. In both cases, the norm of the bivector $|B^{IJ}_\Delta|$ gives the area of the face $\Delta$ and the bivectors satisfy a closure constraint:
\be
\sum_{\Delta\in\pp v}B^{IJ}_\Delta =0.
\ee
For a discussion on how these two sectors correspond to gravity or not, the interested reader can check out for example \cite{plebanski}.

The simplicity constraints come from the fact that all the faces of a given 3-cell $v$ all lay in the same hypersurface. Let us call $\cN_I$ the 4-vector normal to that hypersurface. This leads to very simple constraints on the bivectors $B^{IJ}_\Delta$. We will call them the {\it linear simplicity constraints}. They distinguish the two different sectors:
\be
(s)\quad \cN_I\cdot B^{IJ}_\Delta = 0,\forall \Delta\in\pp v,\qquad\qquad
(\star)\quad \epsilon^{IJ}{}_{KL}\cN_J\cdot B^{KL}_\Delta = 0,\forall \Delta\in\pp v.
\ee
These linear simplicity constraints are at the root of the construction of the  recent EPRL-FK model and other new spinfoam models \cite{EPR,EPRL,FK,LS,sergei}. Here we work {\it without Immirzi parameter}. The Immirzi parameter is taken into account by considering a linear combination of these two sectors, thus leading to a linear simplicity constraint of the type $\cN_I\cdot (B^{IJ}_\Delta+\gamma\epsilon^{IJ}{}_{KL}B^{KL}_\Delta)=0$.

These linear simplicity constraints involve explicitly the time normal $\cN_I$, which is an extra field. This time normal extra-field appears explicitly in some formulations, for instance in the covariant loop gravity framework by Alexandrov and collaborators e.g. \cite{CLQG1,CLQG2}. But it is not usually considered as a fundamental field variable. It is thus convenient for some purposes to reformulate the simplicity constraints solely in term of the $B$-field. This leads to the discrete {\it quadratic simplicity constraints}, which were actually the original formulation of the simplicity constraints:
\be
\forall \Delta,\tDelta\in\pp v,\quad
\epsilon_{IJKL}\,B^{IJ}_\Delta B^{KL}_\tDelta=0.
\ee
Written as such, the  quadratic simplicity constraints truly look like the discretization of the classical simplicity constraints of the continuum  Plebanski theory (e.g. \cite{plebanski}), which turn BF theory into general relativity. The moot point is that these quadratic simplicity constraints do not distinguish the two sectors $(s)$ and $(\star)$ since they are invariant under taking the Hodge dual $B\,\arr\, \star B$. This means that we have to recognize the two sectors by hand when solving these simplicity constraints.

The quantization procedure for the 3-cell is very simple: we associate an irreducible representation of our gauge group $\Spin(4)$ to each face $\Delta$ and we quantize the bivectors $B^{IJ}_\Delta$ as the Lie algebra generators $J^{IJ}_\Delta$ acting in that representation. Then states of the 3-cell $v$ live in the tensor product of the representations of all its faces. Moreover, taking into account the closure constraint,
$$
\sum_{\Delta\in\pp v}B^{IJ}_\Delta =0
\quad\longrightarrow\quad
\sum_{\Delta\in\pp v}J^{IJ}_\Delta =0,
$$
we require that the states associated to the 3-cell must be invariant under the global $\Spin(4)$ action which acts simultaneously on all faces. Thus, the Hilbert space of quantum states of the 3-cell is the space of $\Spin(4)$  intertwiners between the representations attached to its faces.
Since $\Spin(4)\sim\SU_L(2)\times\SU_R(2)$ is the direct product of its two $\SU(2)$ subgroups,
the irreducible representations of $\Spin(4)$  are labeled by a couple of half-integers $(j^L,j^R)$. So we attach a pair of spin to every face $\Delta$ and the intertwiner space for the 3-cell is the tensor product of the space of $\SU_L(2)$ intertwiners between the spins $j^L_\Delta$ and the space of $\SU_R(2)$ intertwiners between the spins $j^R_\Delta$~:
\be
H_{j^L_\Delta,j^R_\Delta}\,\equiv\,
\textrm{Inv}\left[\bigotimes_\Delta V^{j^L_\Delta}\right]\,\otimes\,
\textrm{Inv}\left[\bigotimes_\Delta V^{j^R_\Delta}\right].
\ee
Then we still need to implement the simplicity constraint. We use the quadratic simplicity constraints:
\be
\forall \Delta,\tDelta\in\pp v,\quad
\epsilon_{IJKL}\,J^{IJ}_\Delta J^{KL}_\tDelta
\,=\,0.
\ee
They translate very simply into conditions on the Casimir operators on the intertwiners states. We distinguish the {\it diagonal simplicity constraints} obtained when the faces are the same $\Delta=\tDelta$~,
\be \label{diag}
\forall \Delta,\quad
(\vJ^L_\Delta)^2
\,=\,
(\vJ^R_\Delta)^2,
\ee
and the {\it crossed simplicity constraints} in the case that the two faces are different $\Delta\ne\tDelta$,
\be \label{crosseddiag}
\forall \Delta\ne\tDelta,\quad
(\vJ^L_\Delta\cdot\vJ^L_\tDelta )
\,=\,
(\vJ^R_\Delta\cdot\vJ^R_\tDelta ).
\ee
The diagonal simplicity constraints are simple to impose. We require that the $\Spin(4)$ representation be simple i.e $j^L_\Delta=j^R_\Delta$ for all faces $\Delta$. So we can drop the index $L,R$ for the spin labels and call them simply $j_\Delta=j^L_\Delta=j^R_\Delta$. The crossed simplicity constraints are the true constraints on the intertwiner states and are more complicated to impose.

Naturally, the first approach is to attempt to impose these constraints strongly for looking for intertwiner states $|\psi\ra$ vanishing under the simplicity constraints:
\be
\epsilon_{IJKL}\,J^{IJ}_\Delta J^{KL}_\tDelta\,|\psi\ra=0.
\ee
It was shown (at least in the 4-valent case corresponding to a tetrahedron) in \cite{unique} that these equations have a unique solution once the spins $(j^L_\Delta,j^R_\Delta)$ are specified. This leads to the Barrett-Crane spinfoam model \cite{BC}. This frozen intertwiner issue leads to several problems in the interpretation of the Barrett-Crane model, especially when looking at its relation with the canonical loop quantum gravity framework and when studying the graviton propagator derived in the asymptotical semi-classical regime of the model. This uniqueness of the intertwiner state can be traced back to the fact that the Lie algebra of the quadratic simplicity constraints does not close. Indeed, looking at the crossed simplicity constraints, it is fairly easy to check that:
\be
\left[
\,\vJ^L_{\Delta_1}\cdot\vJ^L_{\Delta_2} \,
-
\,\vJ^R_{\Delta_1}\cdot\vJ^R_{\Delta_2} \,,
\,\vJ^L_{\Delta_1}\cdot\vJ^L_{\Delta_3} \,
-
\,\vJ^R_{\Delta_1}\cdot\vJ^R_{\Delta_3} \,
\right]
\,=\,
\vJ^L_{\Delta_1}\cdot\left(\vJ^L_{\Delta_2}\w \vJ^L_{\Delta_3}\right)
+
\vJ^R_{\Delta_1}\cdot\left(\vJ^R_{\Delta_2}\w \vJ^R_{\Delta_3}\right).
\ee
And so on we generate higher and higher order constraints by computing further commutators. This means that when we impose strongly the quadratic simplicity constraints on the intertwiner state $|\psi\ra$, we are actually also imposing all these higher order constraints. Then it is not surprising to find a unique solution, although it could actually be considered surprising  to find at least one solution.

To remedy this issue, it was proposed to solve the crossed simplicity constraints weakly, either by some coherent state techniques \cite{coh1} or by some Gupta-Bleuler-like method using the linear simplicity constraints \cite{EPR0,EPR,EPRL}. These two approaches were shown to lead to the same spinfoam amplitudes \cite{FK,LS} apart from some subtle cases \cite{engle}. On the one hand, one seeks semi-classical states such that the simplicity constraint is solved in average \cite{coh1,FK},
\be
\la\psi|\vJ^L_{\Delta}\cdot\vJ^L_{\tDelta}|\psi\ra
\,=\,
\la\psi|\vJ^R_{\Delta}\cdot\vJ^R_{\tDelta}|\psi\ra
,
\ee
and minimizing the uncertainty of these operators. On the other hand, one looks for a Hilbert space $H_s[j_\Delta]$ such that the matrix elements of the simplicity constraints on this smaller Hilbert space all vanish \cite{EPR0,LS}:
\be
\forall \phi,\psi\in H_s[j_\Delta],\quad\,
\la\phi|\vJ^L_{\Delta}\cdot\vJ^L_{\tDelta}|\psi\ra
\,=\,
\la\phi|\vJ^R_{\Delta}\cdot\vJ^R_{\tDelta}|\psi\ra.
\ee
At the end of the day, these two methods lead to the same intertwiner space, at least in the present case considered in the Euclidean setting and without Immirzi parameter. We explain this construction in the next section.

\medskip

Our approach in the present paper is to attempt to address these crossed simplicity constraints from the viewpoint of the $\U(N)$ framework for $\SU(2)$ intertwiners. We will propose a closed set of $\U(N)$ simplicity constraints, which can be solved by a straightforward group averaging, and also we will introduce another set of Gupta-Bleuler-like simplicity constraints using the $\U(N)$ creation and annihilation operators.

%%%
\subsection{Using Coherent Intertwiners} \label{SU2coh}
%%%

Let us introduce the coherent intertwiners used to solve weakly the crossed simplicity constraints. First, we need to define the usual $\SU(2)$ coherent states. They are derived by acting with $\SU(2)$ rotations on the highest weight vectors $|j,j\ra$:
\be
\forall g\in\SU(2),\quad
|j,g\ra \,\equiv\, g\,|j,j\ra .
\ee
These states are coherent states \`a la Perelomov and transform consistently under the $SU(2)$ action:
\be
h\,|j,g\ra\,=\,
|j,hg\ra.
\ee
They also satisfy a very simple tensorial property:
\be
|j,g\ra\otimes|\tj,g\ra
\,=\,
g\,(|j,j\ra\otimes |\tj,\tj\ra)
\,=\,
g\,|j+\tj,j+\tj\ra
\,=\,
|j+\tj,g\ra.
\ee
It is easy to compute their expectation values:
\be
\la j,g|\vJ|j,g\ra
\,=\,
j\,\hat{n},
\qquad
\hat{n}=g\,\hat{z}.
\ee
where the unit vector $\hat{n}\in\cS^2$ is obtained by rotating the $z$ axis by the  $\SU(2)$ group element $g$. They are actually coherent states on the 2-sphere $\cS^2\sim \SU(2)/\U(1)$ since $\hat{n}$ only depends on $g$ up to a $\U(1)$ phase. Actually, the standard definition of the $\SU(2)$ coherent states involves a choice of section and we usually choose the unique group element $g(\hat{n})$ for a given unit vector $\hat{n}$ such that the rotation axis lays in the $(xy)$ plane. Then the coherent state is defined as $|j,\hat{n}\ra \,\equiv\, g(\hat{n})\,|j,j\ra=|j,g(\hat{n})\ra$.
Finally, these states minimize the uncertainty relation:
\be
\la j,g|\vJ^2|j,g\ra-\la j,g|\vJ|j,g\ra^2
\,=\,
j(j+1)-j^2
\,=\,
j.
\ee

Then $N$-valent coherent intertwiners are defined following \cite{coh1} by tensoring $N$ such $\SU(2)$ coherent states and group averaging this tensor product in order to get an intertwiner. More precisely, we choose $N$ representations of $\SU(2)$ labeled by the spins $j_1,..,j_N$ and $N$ unit 3-vectors $\hat{n}_1,..,\hat{n}_N$, then we define:
\be
||j_i,\hat{n}_i\ra
\,\equiv\,
\int_{\SU(2)} dg\, g\triangleright \bigotimes_{i=1}^N
|j_i,\hat{n}_i\ra
\,=\,
\int_{\SU(2)} dg\,\bigotimes_{i=1}^N
gg(\hat{n}_i)\,|j_i,j_i\ra.
\ee

\medskip

Now coming back to the $\Spin(4)\sim\SU_L(2)\times\SU_R(2)$  and the simplicity constraints,
we use simple $\Spin(4)$ representations with $j^L_i=j^R_i$ and double the labels of the coherent states introducing unit 3-vectors $\hat{n}^{L,R}_1,..,\hat{n}^{L,R}_N$. Considering the tensor product of $\SU(2)$ coherent states $|j,\hat{n}^L\ra\otimes|j,\hat{n}^R\ra$, the expectation values of the $\spin(4)$ generators $\vJ_i^{L,R}$ are $j\,\hat{n}^{L,R}$. These two 3-vectors with equal norm define a single bivector $B\in\w^2\R^4$ being its self-dual and anti-self dual components. Then considering $N$ such coherent states $|j_i,\hat{n}_i^L\ra\otimes|j_i,\hat{n}_i^R\ra$, their expectation values give $N$ bivectors $B_i$. We now impose the simplicity constraints on these classical bivectors .

\medskip

First looking  at the sector $(s)$, the fact that the bivectors $B_i$ all share a same   time normal $\cN$ translates to the existence of a $\SU(2)$ transformation mapping simultaneously all the self-dual part  (left) onto the anti-self-dual part (right). Moreover, this $\SU(2)$ group element defines uniquely the normal 4-vector~:
\be
\forall i,\, \cN\cdot B_i=0
\qquad\Leftrightarrow\qquad
\exists g\in\SU(2),\quad \forall i,\,\hat{n}_i^L=g\,\hat{n}_i^R
,\quad
\cN=(g,\id)\triangleright \cN^{(0)},
\ee
where $\cN^{(0)}=(1,0,0,0)$ is the  reference unit time vector and the $\Spin(4)$ group element $(g,\id)$ is defined through its left/right factors.
Imposing this constraint on the labels of the coherent states, we define a $\Spin(4)$ coherent intertwiner by group averaging. This coherent intertwiner is labeled by the representation label, plus the unit 3-vectors $\hat{n}_i^R$, plus the group element $g$ which gives the time normal and the rotation which  defines the components $\hat{n}_i^L$ from the $\hat{n}_i^R$~:
\bes
||j_i,\hat{n}_i,g\ra_s
&=&
\int_{\SU_L(2)\times\SU_R(2)} dg_Ldg_R\,
\left[g_L\triangleright \bigotimes_{i=1}^N g|j_i,\hat{n}_i\ra\right]
\otimes \left[g_R\triangleright\bigotimes_{i=1}^N |j_i,\hat{n}_i\ra\right]\nn\\
&=&
\left[\int_{\SU(2)} dg_L\,g_L\triangleright \bigotimes_{i=1}^N |j_i,\hat{n}_i\ra\right]
\otimes \left[\int_{\SU(2)} dg_R\,g_R\triangleright\bigotimes_{i=1}^N |j_i,\hat{n}_i\ra\right].
\ees
Two things are obvious from this expression:
\begin{itemize}
\item The $\Spin(4)$ group averaging erases the group element $g$ and thus all the data about the precise time normal $\cN$. In particular, we can drop the label $g$ and call these states simply $||j_i,\hat{n}_i\ra_s$.
\item This states $||j_i,\hat{n}_i\ra_s$ are the tensor product of two identical $\SU(2)$ intertwiners for the left and right parts. In particular, they obvious satisfy the quadratic simplicity constraints:
    \be
    \forall i,j,\quad
    \la\vJ^L_i\cdot\vJ^L_j\ra
    \,=\,
    \la\vJ^R_i\cdot\vJ^R_j\ra.
    \ee
    Moreover, they minimize the uncertainty by definition.
\end{itemize}
We can go one step further by re-writing these states,
\be
||j_i,\hat{n}_i\ra_s
\,=\,
\int_{\Spin(4)}dG\,G\triangleright \bigotimes_{i=1}^N |j_i,\hat{n}_i\ra_L\otimes|j_i,\hat{n}_i\ra_R
\,=\,
\int_{\Spin(4)}dG\,G\triangleright \bigotimes_{i=1}^N |2j_i,\hat{n}_i\ra,
\ee
where we used the tensorial property of the $\SU(2)$ coherent states $|j,\hat{n}\ra\otimes|\tj,\hat{n}\ra=|j+\tj,\hat{n}\ra$. This shows that the coherent states are exactly the EPR states \cite{EPR0,EPR} which form a Hilbert space $H_s[j_\Delta]$ solving weakly the simplicity constraints \cite{FK,LS}.

\medskip

Second, we consider the dual sector $(\star)$ and we follow the same procedure. The only difference is a sign in solving the corresponding linear simplicity constraints~:
\be
\forall i,\, \eps\cN B_i=0
\qquad\Leftrightarrow\qquad
\exists g\in\SU(2),\quad \forall i,\,\hat{n}_i^L=-g\,\hat{n}_i^R.
\ee
This leads to similar coherent states \cite{FK,LS}:
\bes
||j_i,\hat{n}_i\ra_\star
&=&
\left[\int_{\SU(2)} dg_L\,g_L\triangleright \bigotimes_{i=1}^N |j_i,-\hat{n}_i\ra\right]
\otimes \left[\int_{\SU(2)} dg_R\,g_R\triangleright\bigotimes_{i=1}^N |j_i,\hat{n}_i\ra\right],\\
&=&
\left[\int_{\SU(2)} dg_L\,g_L\triangleright \bigotimes_{i=1}^N \overline{|j_i,\hat{n}_i\ra}\right]
\otimes \left[\int_{\SU(2)} dg_R\,g_R\triangleright\bigotimes_{i=1}^N |j_i,\hat{n}_i\ra\right].
\ees
Once again, it is obvious to check that these states satisfy the quadratic simplicity constraints in expectation value. Also, the information about the time normal is completely erased by the group averaging. Finally, the key difference with the previous sector $(s)$ is that these intertwiner states generate the whole intertwiner space and do not form a subspace. This ansatz looks more like a fuzzy version of the Barrett-Crane intertwiner.

\medskip

This concludes our quick overview of the standard way to deal with the crossed simplicity constraints. We will now review the $\U(N)$ framework for intertwiners and see what this new approach can say about the simplicity constraints.

%%%%%%%%%
\section{The $\U(N)$ Framework for Intertwiners}
%%%%%%%%%

%%%
\subsection{Review of the $\U(N)$ operators}
%%%

%\begin{itemize}
%
%\item Initial goal: find a closed alg of invariant operators to replace the alg of scalar product operators
%
%\item The $\U(N)$ finally goes beyond this initial objective, it offers a whole new perspective on the intertwiner space. Introduce $\u(N)$ operators $E_{ij}$ and annihiliation/creation operators $F,F\dag$
%
%\item Define $\U(N)$ coherent states, give the expectation values of the $E$ ops and discuss their geometric interpretation
%
%\end{itemize}

The standard invariant operators considered on the space of $\SU(2)$ intertwiner  are the scalar product operators. They act on pairs of legs $(i,j)$ of the intertwiner and are simply $\vJ_i\cdot\vJ_j$. As presented above, the discrete simplicity constraints are usually formulated in term of these operators. An important issue is that the algebra of the scalar product operators does not close. More precisely, the commutator of two scalar product operators gives a operator of order 3 in the $\vJ$'s:
\be
[ \vec{J}_i\cdot \vec{J}_j\,, \, \vec{J}_i\cdot \vec{J}_k]=\,i\vec{J}_i\cdot(\vec{J}_j\wedge \vec{J}_k),
\ee
and so on generating higher and higher order operators. This leads directly to problems when one tries to define coherent intertwiner states minimizing the corresponding uncertainty relations or when one attempts to solve constraints such as the simplicity constraints.  The approach followed in \cite{UN1} was to use Schwinger's representation of the $\su(2)$ algebra in term of harmonic oscillators to identify a new family of invariant operators, whose Lie algebra closes and which would still generate the full algebra of invariant operators acting on the intertwiner space. This leads to the $\U(N)$ framework for $\SU(2)$ intertwiners \cite{UN1,UN2,UN3}, which actually goes beyond this initial objective and offers a whole new perspective on the intertwiner space. It shows that the intertwiner space carries a natural representation of the $\U(N)$ unitary group and allows to build semi-classical coherent states transforming consistently under the $\U(N)$ action. It also uncovers a deep relation between the intertwiner space and the Grassmannian spaces, which could prove very useful to understand the geometry of the intertwiner space and its (semi-)classical interpretation. We review this formalism below.

\medskip

Let us consider the Hilbert spaces of intertwiners between $N$
irreducible $\SU(2)$-representations of spin $j_1,..,j_N$~:
\be
\cH_{j_1,..,j_N}\,\equiv\, \textrm{Inv}[V^{j_1}\otimes..\otimes V^{j_N}].
\ee
We further introduce the space of intertwiners with $N$ legs and fixed total area $J=\sum_i j_i$~:
\be
\cH_N^{(J)}\,\equiv\,\bigoplus_{\sum_i j_i=J}\cH_{j_1,..,j_N},
\ee
and the full Hilbert space of $N$-valent intertwiners:
\be
\cH_N\,\equiv\,\bigoplus_{\{j_i\}}\cH_{j_1,..,j_N}\,=\, \bigoplus_{J\in\N}\cH_N^{(J)}.
\ee
The key result of the $\U(N)$ formalism is that there is a natural action of $\U(N)$ on the intertwiner space $\cH_N$ \cite{UN1}. More precisely the intertwiner spaces $\cH_N^{(J)}$
carry irreducible representations of $\U(N)$ \cite{UN2}. Finally the full space $\cH_N$ can be endowed with a Fock space structure with creation and annihilation operators compatible with
the $\U(N)$ action \cite{UN3}.

This $\U(N)$ formalism is based on the Schwinger representation of the $\su(2)$ Lie algebra in term of harmonic oscillators. Let us introduce $2N$ oscillators with creation operators $a_i,b_i$ with
$i$ running from 1 to $N$:
$$
[a_i,a\dag_j]=[b_i,b\dag_j]=\delta_{ij}\,,\qquad [a_i,b_j]=0.
$$
The generators of the $\SU(2)$ transformations acting on each leg of
the intertwiner are realized as quadratic operators in term of the
oscillators:
\be
J^z_i=\f12(a\dag_i a_i-b\dag_ib_i),\qquad
J^+_i=a\dag_i b_i,\qquad
J^-_i=a_i b\dag_i,\qquad
E_i=(a\dag_i a_i+b\dag_ib_i).
\ee
The $J_i$'s satisfy the standard commutation algebra while the total
energy $E_i$  is a Casimir operator:
\be
[J^z_i,J^\pm_i]=\pm J^\pm_i,\qquad
[J^+_i,J^-_i]=2J^z_i,\qquad
[E_i,\vec{J}_i]=0.
\ee
The correspondence with the standard $|j,m\ra$ basis of $\su(2)$
representations is simple:
\be
|n_a,n_b\ra_{OH}=|\f12(n_a+n_b),\f12(n_a-n_b)\ra\,,\qquad
|j,m\ra=|j+m,j-m\ra_{OH}
\ee
where $m$ is the eigenvalue of $J^z$ defined as
the half-difference of the energies between the two oscillators,
while the total energy $E_i$ gives twice the spin, $2j_i$, living on
the $i$-th leg of the intertwiner.

Intertwiner states are by definition invariant under the global
$\SU(2)$ action, generated by:
\be
J^z=\sum_{i=1}^N J^z_i,\qquad
J^\pm=\sum_i J^\pm_i.
\ee
Then operators acting on the intertwiner space need to commute with
these operators too. The simplest family of invariant operators was
identified in \cite{UN1} and are quadratic operators acting on
couples of legs:
\be
E_{ij}=a\dag_ia_j+b\dag_ib_j, \qquad
E_{ij}\dag=E_{ji}.
\ee
The main result is that these operators are invariant under
global $\SU(2)$ transformations and form a $\u(N)$ algebra:
\be
[\vec{J},E_{ij}]=0,\qquad
[E_{ij},E_{kl}]\,=\,
\delta_{jk}E_{il}-\delta_{il}E_{kj}.
\ee
The diagonal operators $E_i\equiv E_{ii}$ form the Cartan
sub-algebra of $\u(N)$, while the off-diagonal operators $E_{ij}$
with $i\ne j$ are the raising and lowering operators. As said
earlier, the generators $E_i$ give twice the spin $2j_i$ while the
$\U(1)$ Casimir $E=\sum_i E_i$ will give twice the total area,
$2J\equiv \sum_i 2j_i$. Then all operators $E_{ij}$ commute with the
$\U(1)$ Casimir, thus leaving the total area $J$ invariant:
\be
[E_{ij},E]=0.
\ee
The usual $\SU(2)$ Casimir operators have a simple expression in
term of these $\u(N)$ generators:
\be
(\vec{J}_i)^2=\f{1}{2}E_i\left(\f{E_i}{2}+1\right),\qquad
\forall i\ne j,\,\,
(\vec{J}_i\cdot \vec{J}_j)
\,=\,
\f12E_{ij} E_{ji} -\f14E_iE_j-\f12 E_i.
%\,=\,
%\f12E_{ji} E_{ij} -\f14E_iE_j-\f12 E_j.
\label{scalaropE}
\ee
Let us point out that case $i=j$ of $(\vec{J}_i\cdot
\vec{J}_j)$ does not give back exactly the formula for
$(\vec{J}_i)^2$ due to the ordering of the oscillator operators. The
two formula agree on the leading order quadratic in $E_i$ but
disagree on the correction linear in $E_i$.

The next point is that explicit definition of the $E_{ij}$'s in term of
harmonic oscillators leads to quadratic constraints on these
operators as shown in \cite{UN2}~:
\be
\forall i,\quad
\sum_j E_{ij}E_{ji}=E_{i} \left(\f E2+N-2\right),
\ee
where we have assumed that the global $\SU(2)$ generators $\vec{J}$
vanish.
These quadratic constraints on the $E_{ij}$ operators lead to non-trivial restrictions on the representations of $\u(N)$ obtained from this construction. To solve them, the method
used in \cite{UN2} is to apply them to a highest weight vector.
This allows to identify the representations corresponding to the intertwiner spaces $\cH_N^{(J)}$ at fixed total area $J=\sum_i j_i$. The highest weight vector $v_N^{(J)}$ diagonalizes the generators of the Cartan sub-algebra $E_i$ and vanishes under the action of the raising operators $E_{ij}\,v=0$ for all $i<j$. The $N$ eigenvalues depend very simply on the area $J$~:
\be
E_1\,v_N^{(J)}=E_2\,v_N^{(J)}=J\,v_N^{(J)},\qquad
E_k\,v_N^{(J)}=0,\,\forall k\ge 3.
\ee
This highest weight vector corresponds to a bivalent intertwiner between two legs  both carrying the spin $\f J2$.
One can compute the corresponding value of the quadratic $\U(N)$ Casimir using the previous quadratic constraints:
\be
\sum_{i,j} E_{ij}E_{ji}=E\left(\f E2+N-2\right)= 2J(J+N-2),
\ee
and the dimension of $\cH_N^{(J)}$ in term of binomial coefficients using the standard formula for $\U(N)$ representations:
\be
D_{N,J}\,\equiv\,
\dim \cH^{(J)}_N
\,=\,
\f{1}{J+1}\bin{J}{N+J-1}\bin{J}{N+J-2}
\,=\,
\frac{(N+J-1)!(N+J-2)!}{J!(J+1)! (N-1)!(N-2)!}\,.
\label{dimNJ}
\ee

\medskip

Next, we introduce annihilation and creation operators to move
between the spaces $\cH_N^{(J)}$ with different areas $J$ within the
bigger Hilbert space of all intertwiners with $N$ legs \cite{UN3}.
We define the new operators:
\be
F_{ij}=(a_i b_j - a_j b_i),\qquad
F_{ji}=-F_{ij}.
\ee
These are still invariant under global $\SU(2)$ transformations, but
they do not commute anymore with the total area operator $E$. They nevertheless form a closed algebra together with the operators $E_{ij}$:
\bea
{[}E_{ij},E_{kl}]&=&
\delta_{jk}E_{il}-\delta_{il}E_{kj}\nn\\
{[}E_{ij},F_{kl}] &=& \delta_{il}F_{jk}-\delta_{ik}F_{jl},\qquad
{[}E_{ij},F_{kl}\dag] = \delta_{jk}F_{il}\dag-\delta_{jl}F_{ik}\dag, \\
{[} F_{ij},F\dag_{kl}] &=& \delta_{ik}E_{lj}-\delta_{il}E_{kj} -\delta_{jk}E_{li}+\delta_{jl}E_{ki}
+2(\delta_{ik}\delta_{jl}-\delta_{il}\delta_{jk}), \nn\\
{[} F_{ij},F_{kl}] &=& 0,\qquad {[} F_{ij}\dag,F_{kl}\dag] =0.\nn
\eea
The annihilation operators $F_{ij}$ allow to go from $\cH_N^{(J)}$ to
$\cH_N^{(J-1)}$ while the creation operators $F\dag_{ij}$ raise the
area and go from $\cH_N^{(J)}$ to $\cH_N^{(J+1)}$.
We can re-express the scalar product operators in term of this new set of operators:
\bes
\label{scalaropF}
\vec{J}_i\cdot \vec{J}_j&=& -\f12 F_{ij}^\dagger F_{ij} +\f14 E_iE_j,  \\
&=& -\f12 F_{ij}F_{ij}^\dagger +\f14 (E_i+2)(E_j+2). \nn
\ees
%\be \label{JF}
% \left \{\tabl{l}{
%\vec{J}_i\cdot \vec{J}_j= -\f12 F_{ij}^\dagger F_{ij} +\f14 E_iE_j  \\
%\vec{J}_i\cdot \vec{J}_j= -\f12 F_{ij}F_{ij}^\dagger +\f14 (E_i+2)(E_j+2)  \\
%} \right.
%\ee
This formula is explicitly symmetric in the edge labels $i \leftrightarrow j$ contrary to the previous formula \Ref{scalaropE} in terms of the $E_{ij}$-operators.
Finally, as shown in \cite{UN3} and as we review below, these operators can be used to construct coherent states transforming consistently under $\U(N)$ transformations. These $\U(N)$ coherent states will turn out very convenient in order to re-express and solve the simplicity constraints.

%These new operators satisfy new quadratic constraints together with
%the $E_{ij}$.  Starting with their explicit definitions in  term of
%oscillators, it is easy to get on the intertwiner space (assuming global $\SU(2)$ invariance, $\vec{J}=0$):
%%\beq
%%\sum_k F\dag_{ik}E_{jk} &=& F\dag_{ij}\, \frac{E}{2} \,,\\
%%\sum_k E_{jk} F\dag_{ik}     &=& F\dag_{ij}\left(\frac{E}{2}+N-1\right) \,,\\
%%\sum_k F\dag_{ik}F_{jk} &=& E_{ij}
%%\left(\frac{E}{2}+1\right)\,\,.
%%\eeq
%\beq
%&&\sum_k F\dag_{ik}E_{jk} = F\dag_{ij}\, \frac{E}{2}, \qquad\qquad\quad
%\sum_k E_{jk} F\dag_{ik} = F\dag_{ij}\left(\frac{E}{2}+N-1\right),\label{constraint1}\\
%&&\sum_k E_{kj}F_{ik} = F_{ij}\, \left(\frac{E}{2}-1\right), \qquad
%\sum_k F_{ik} E_{kj}  = F_{ij}\left(\frac{E}{2}+N-2\right),\label{constraint2}\\
%&&\sum_k F\dag_{ik}F_{jk} = E_{ij}
%\left(\frac{E}{2}+1\right),\qquad
%\sum_k F_{jk}F\dag_{ik} = (E_{ij}+2\delta_{ij})
%\left(\frac{E}{2}+N-1\right)\,.\label{constraint3}
%\eeq

%%%
%\subsection{Reviewing Coherent Intertwiners}
\subsection{Revisiting Coherent Intertwiners}
%%%

%\begin{itemize}
%
%\item Express coherent intertwiners in term of the oscillator operators
%
%\item Compare them with the $\U(N)$ coherent states
%
%\end{itemize}

To define coherent intertwiner states, we attach a spinor $z_i$ to each leg of the intertwiner:
$$
z_i=\mat{c}{z^0_i\\z^1_i}.
$$
Basically, the first component $z^0_i$ is the label of the coherent state for the oscillator $a_i$ while the second component $z^1_i$ corresponds to the oscillator $b_i$.
Let us first clear up the geometrical meaning of the spinors $z_i$. Considering a spinor $z$, the matrix $|z\ra\la z|$ is a $2\times 2$ matrix which can be decomposed on the Pauli matrices $\sigma_a$ (taken Hermitian and normalized so that $(\sigma_a)^2=\id$). This defines a 3-vector $\vec{V}(z)$:
\be \label{vecV}
|z\ra \la z| = \f12 \left( {\la z|z\ra}\id  + \vec{V}(z)\cdot\vec{\sigma}\right),
\ee
where the norm of the vector is $|\vec{V}(z)| = \la z|z\ra= |z^0|^2+|z^1|^2$ and its components are given explicitly as~\footnotemark:
\be
V^z=|z^0|^2-|z^1|^2,\qquad
%V^+=2z^0\bar{z}^1,\qquad
V^x=2\,\Re\,(\bar{z}^0z^1),\qquad
%J^x=2{\rm Re}\,(\zo\bzi),\qquad
V^y=2\,\Im\,(\bar{z}^0z^1).
%J^y=2{\rm Im}\,(\zo\bzi),\qquad
\ee
\footnotetext{
Remember the convention for the $\pm$ generators:
$$
\sigma_\pm=\sigma_x\pm i\sigma_y,\quad
\sigma_x=\f12(\sigma_++\sigma_-), \quad
\sigma_y=-i\f12(\sigma_+-\sigma_-).
%,\quad
%V^\pm=\f12(V^x\mp iV^y).
$$
}
The spinor $z$ is entirely determined by the corresponding 3-vector $\vec{V}(z)$ up to a global phase. Following \cite{UN3}, we also introduce the map $\varsigma$ between spinors:
\be
\varsigma\begin{pmatrix}z^0\\ z^1\end{pmatrix}
\,=\,
\begin{pmatrix}-\bar{z}^1\\\bar{z}^0 \end{pmatrix},
\qquad \varsigma^{2}=-1.
\ee
This is an anti-unitary map, $\la \varsigma z| \varsigma w\ra= \la w| z\ra=\overline{\la z| w\ra}$, and we will write the related state as
$$
|z]\equiv \varsigma  | z\ra.
$$
This map $\varsigma$ maps the 3-vector $\vec{V}(z)$ onto its opposite:
\be
|z][  z| = \f12 \left({\la z|z\ra}\id - \vec{V}(z)\cdot\vec{\sigma}\right).
\ee
Finally coming back to the intertwiner with $N$ legs, we have $N$ spinors and their corresponding 3-vectors $\vV(z_i)$. Typically, we can require that the $N$ spinors satisfy a closure constraint, $\sum_i \vec{V}(z_i)=0$. This can be written in term of $2\times 2$ matrices:
\be
\sum_i |z_i\ra \la z_i|=A(z)\id,
\qquad\textrm{with}\quad
A(z)\equiv\f12\sum_i \la z_i|z_i\ra=\f12\sum_i|\vec{V}(z_i)|.
\ee
It translates into quadratic constraints on the spinors:
\be
\sum_i z^0_i\,\bar{z}^1_i=0,\quad
\sum_i \left|z^0_i\right|^2=\sum_i \left|z^1_i\right|^2=A(z),
\label{closure}
\ee
which means that the two components of the spinors, $z^0_i$ and $z^1_i$, are orthogonal $N$-vectors of equal norm.

\medskip

Then we can define coherent intertwiner states \cite{coh1,coh2,coh3}. First, for a given leg, we define the  $\SU(2)$ coherent states labeled by the spin $j_i$ and the spinor $z_i$:
\be
|j_i,z_i\ra\,\equiv\,
\f{(z^0_ia\dag_i+z^1_ib\dag_i)^{2j_i}}{\sqrt{(2j_i)!}}\,|0\ra.
\ee
This vector clearly lives in the irreducible $\SU(2)$-representation of spin $j_i$ since it's an eigenvector of the energy $E_i$ with value $2j_i$. To show that it transforms coherently under $\SU(2)$, we compute the  $\SU(2)$ action. Dropping the index $i$, $\SU(2)$ rotations are parameterized by an angle $\theta$ and a unit 3-vector $\hat{v}\in\cS_2$:
\be
g(\theta,\hat{v})
\,\equiv\,
e^{i\theta \hat{v}\cdot\vec{J}}
\,=\,
e^{i\theta (v_zJ_z+\f{\bv}2J_++\f{\bar{\bv}}2J_-)},
\qquad
|\vec{v}|^2=v_z^2+|\bv|^2=1,\qquad
v_z=\cos\phi,\quad \bv=e^{i\psi}\sin\phi.
\ee
It is represented by a $2\times 2$ matrix in the fundamental spin-$\f12$ representation:
\be
g(\theta,\hat{v})
\,=\,
e^{i\theta \hat{v}\cdot\f{\vec{\sigma}}2}
\,=\,
\mat{cc}{\cos\f\theta2+i\cos\phi\sin\f\theta2 & i e^{i\psi}\sin\phi\sin\f\theta2 \\
i e^{-i\psi}\sin\phi\sin\f\theta2 & \cos\f\theta2-i\cos\phi\sin\f\theta2}
\,\in\SU(2).
\ee
To compute the action of $\SU(2)$, we first compute the following commutator:
\be
\left[\vec{v}\cdot \vJ, (z^0a\dag+z^1b\dag)\right]=
\left((\vec{v}\cdot\f{\vec{\sigma}}2)\, z\right)^0a\dag+\left((\vec{v}\cdot\f{\vec{\sigma}}2)\, z\right)^1b\dag,
\ee
which gets easily exponentiated:
\be
e^{i\theta \hat{v}\cdot\vec{J}}\,(z^0a\dag+z^1b\dag)\,e^{-i\theta \hat{v}\cdot\vec{J}}
=e^{[i\theta \hat{v}\cdot\vec{J},\cdot]}\,(z^0a\dag+z^1b\dag)
\,=\,
(\tz^0a\dag+\tz^1b\dag),
\qquad\textrm{with}\quad
\tz=e^{i\theta \hat{v}\cdot\f{\vec{\sigma}}2}\,z=g(\theta,\hat{v})\,z.
\ee
This shows that the states introduced above are proper $\SU(2)$ coherent states:
\be
g(\theta,\hat{v})\,|j,z\ra
\,=\,
|j,\,g(\theta,\hat{v})\,z\ra
\ee
This means that these are the standard $\SU(2)$ coherent states \`a la Perelomov. Indeed, one can always set $\tz^1$ to 0, or reversely get any arbitrary state from the initial state without any $b$-excitation. Such an initial state actually corresponds to the highest weight vector $|j,j\ra$ of the $\SU(2)$-representation of spin $j$. More precisely, we act on that highest weight vector with a $\SU(2)$ transformation parameterized by $\alpha$ and $\beta$ satisfying $|\alpha|^2+|\beta|^2=1$:
\be
|j,j\ra=|2j,0\ra_{OH}=\f{(a\dag)^{2j}}{\sqrt{(2j)!}}\,|0\ra\,\arr\,
\mat{cc}{\alpha &\beta \\ -\bar{\beta}&\bar{\alpha}}\,|j,j\ra
\,=\,
|j,\,\mat{c}{\alpha \\ -\bar{\beta}}\ra,
\ee
\be
|j,z\ra=\,\la z|z\ra^j
\mat{cc}{\f{z^0}{\sqrt{\la z|z\ra}} &\f{-\bar{z}^1}{\sqrt{\la z|z\ra}}  \\ \f{z^1}{\sqrt{\la z|z\ra}} &\f{\bar{z}^0}{\sqrt{\la z|z\ra}} }\,|j,j\ra.
\ee
%\be
%|j,j\ra=|2j,0\ra_{OH}=\f{(a\dag)^{2j}}{\sqrt{(2j)!}}\,|0\ra\,\arr\,
%\mat{cc}{\alpha &\beta \\ -\bar{\beta}&\bar{\alpha}}\,|j,j\ra
%\,=\,
%|j,\,\mat{c}{\alpha \\ -\bar{\beta}}\ra,\qquad
%|j,z\ra=\,\la z|z\ra^j
%\mat{cc}{\f{z^0}{\sqrt{\la z|z\ra}} &\f{-\bar{z}^1}{\sqrt{\la z|z\ra}}  \\ \f{z^1}{\sqrt{\la z|z\ra}} &\f{\bar{z}^0}{\sqrt{\la z|z\ra}} }\,|j,j\ra.
%\ee
%
We also give the scalar product between two such $\SU(2)$ coherent states:
\be
\la j,w|j,z\ra=\la w|z\ra^{2j},
\ee
and the expectation values of the $\su(2)$ generators:
\be
\la J_z \ra\,\equiv\,
\f{\la j,z|J_z|j,z\ra}{\la j,z|j,z\ra}
\,=\,
j\,\f{|z^0|^2-|z^1|^2}{|z^0|^2+|z^1|^2},
\quad
\la J_+ \ra\,=\,
2j\,\f{\bar{z}^0z^1}{|z^0|^2+|z^1|^2},
\quad\Rightarrow\quad
\la \vJ \ra\,=\,
\,=\,
j\,\f{\vV}{|\vV|},
\ee
as expected.
Finally, expanding these states explicitly on the standard basis for harmonic oscillators,
$$
|j,z\ra\,=\,\sum_{n=0}^{2j}\sqrt{\bin{n}{2j}}\,(z^0)^n(z^1)^{2j-n}\,|n,2j-n\ra_{OH},
$$
and following the usual calculation done with oscillator coherent states (as shown in appendix \ref{cohHO}), we can decompose the identity on the Hilbert space $V^j$ in term of these $\SU(2)$ coherent states:
\be
\id_j \,=\,
\sum_{n=0}^{2j} |n,2j-n\ra_{OH}\,{}_{OH}\la n,2j-n|
\,=\,
\f{1}{(2j)!}\int [d^2 z^0d^2z^1]\,\f{e^{-\la z|z\ra}}{\pi^2}\,|j,z\ra\la j,z|.
\ee
One can check that taking the trace of this expression and using the formula for the scalar product between coherent states give back as expected $\tr \,\id_j=(2j+1)$. Let us emphasize a last point that the projector $|j,z\ra\la j,z|$ does not depend on the overall phase of the spinor $z$ but only on the corresponding 3-vector $\vV(z)$.

Coherent intertwiners are then defined following \cite{coh1} by group averaging the tensor product of $\SU(2)$ coherent states attached to every leg of the intertwiner:
\be
||\{j_i,z_i\}\ra
%||\{j_i\},\{z_i\}\ra
\,\equiv\,\int_{\SU{2}}dg\, g\rhd\bigotimes_{i=1}^N|j_i,z_i\ra.
\ee
These are the standard coherent intertwiners used in the construction of the EPRL-FK spinfoam models and their boundary states \cite{FK,LS}. Following the logic of \cite{coh1}, we can write the identity on the intertwiner space $\cH_{j_1,..,j_N}$ in term of these coherent intertwiners:
\be
\label{idcohint}
\id_{\cH_{j_1,..,j_N}}
\,=\,
\int \prod_i \f{e^{-\la z_i|z_i\ra}\,d^4z_i}{(2j_i)!\pi^2}\,
||\{j_i,z_i\}\ra\la\{j_i,z_i\}||,
%\f{1}{\prod_i (2j_i)!}
%\int \prod_i \f{e^{-\la z_i|z_i\ra}\,d^4z_i}{\pi^2}\,
%||\{j_i,z_i\}\ra\la\{j_i,z_i\}||,
\ee
where we used the fact that the spinor norm $\la z|z\ra$ is invariant under the $\SU(2)$ action. Finally, the main point shown in \cite{coh1} is that this integral is dominated by intertwiners satisfying the closure constraint $\sum_i j_i \vV(z_i)/|\vV(z_i)|=0$ while the norm of intertwiners not satisfying this closure constraint is exponentially suppressed. It is also possible to write a decomposition of the identity on the intertwiner space restricting the integral to intertwiners satisfying exactly the closure constraint by modifying slightly the measure \cite{coh2,coh3}. This is achieved through considering and gauging out the $\SL(2,\C)$ action on spinors complexifying the previous $\SU(2)$ action.

\medskip

We are now ready to define the $\U(N)$ coherent states. Their definition is slightly more complicated. Following \cite{UN3}, we define the following antisymmetric matrix $Z_{ij}$, which is holomorphic in the spinors $z_i$  and anti-symmetric in $i\leftrightarrow j$~:
\be
Z_{ij}\,\equiv\,
[z_i|z_j\ra
\,=\,
(z^0_iz^1_j-z^0_jz^1_i),
\ee
and the corresponding creation operator:
\be
F\dag_Z
\,\equiv\,
\f12\sum Z_{ij} F\dag_{ij}
=\f12\sum (z^0_iz^1_j-z^0_jz^1_i)\, F\dag_{ij},
\ee
which is also holomorphic in $z$.
A crucial property of this matrix $Z$ is the Pl\"ucker relation $Z_{ik} Z_{jl}-Z_{il} Z_{jk}= Z_{ij} Z_{kl}$ which holds for any indices $i,j,k,l$.
The $\U(N)$ coherent states are then labeled by the total area $J$ and the $N$ spinors $z_i$:
\be
|J,\{z_i\}\ra\,\equiv\,
\f{1}{\sqrt{J!(J+1)!}}\,(F\dag_Z)^J\,|0\ra.
\ee
This state clearly lives in $\cH_N^{(J)}$ since it is an intertwiner (invariant under global $\SU(2)$ transformation) and is an eigenvector of the total area operator $E$ with value $2J$.
Notice that the behavior under rescaling of this coherent state is very simple:
\be
z_i\arr \lambda z_i,\quad
Z_{ij}\arr \lambda^2 Z_{ij},\qquad
|J,\{\lambda z_i\}\ra =\lambda^{2J}\,|J,\{z_i\}\ra.
\ee

Now we assume that the spinors $z_i$ satisfy exactly the closure condition $\sum_i \vV(z_i)=0$ introduced earlier in \Ref{closure}. We can compute the norm of these states:
\be
\la J,\{z_i\}|J,\{z_i\}\ra=(A(z))^{2J}
\,=\,
\left(\f12\sum_i \la z_i|z_i\ra\right)^{2J}
\,=\,
\left(\f12\sum_i |\vec{V}(z_i)|\right)^{2J}.
\ee
Then, as shown in \cite{UN3}, these states are coherent under the action of $\U(N)$:
\be
\forall u\in\U(N),\quad
\hat{u}\,|J,\{z_i\}\ra
\,=\,
|J,\{(uz)_i\}\ra,
\ee
where $\hat{u}$ is the operator representing the unitary transformation $u$, that is for an arbitrary anti-Hermitian matrix $\alpha$~:
\be
u=e^\alpha \quad\rightarrow\quad
\hat{u}\equiv e^{E_\alpha}\equiv e^{\sum_{ij} \alpha_{ij}E_{ij}}.
\ee
The $\U(N)$ action on the $N$ spinors  is the natural one:
\be
z_i\,\rightarrow\, (uz)_i=\sum_j u_{ij}z_j,\qquad
z\arr uz,\quad Z\arr uZu^t.
\ee
This $\U(N)$-action is proved by computing explicitly the action of $\hat{u}$ on the $F\dag$-operators \cite{UN3}:
\be
[E_\alpha,F\dag_Z]=F\dag_{\alpha Z + Z\alpha^t}\quad\Rightarrow\quad
e^{E_\alpha}F\dag_Ze^{-E_\alpha}=F\dag_{e^\alpha Z e^{\alpha^t}}.
\ee
Here is a summary of the properties  of these $\U(N)$ coherent states already proved in \cite{UN3}:
\begin{itemize}
\item They transform simply under $\U(N)$-transformations:
$u\,|J,\{z_i\}\ra\,=\,|J,\{(u\,z)_i\}\ra$. This key property actually holds also if the spinors do not satisfy the closure condition.

\item They are coherent states {\it \`a la} Perelomov  and are obtained by the action of $\U(N)$ on highest weight states. These highest weight vectors correspond to bivalent intertwiners such as the state defined by the spinors $z_1=(z^0,z^1)$, $z_2=\varsigma z_1=(-\bar{z}^1,\bar{z}^0)$ and $z_k=0$ for ${k\ge 3}$. This only holds if one assumes that the spinors satisfy the closure constraint. Indeed, $\U(N)$ transformations conserve the closure condition and the spinors defining the bivalent intertwiner satisfy it.

\item For large areas $J$, they are semi-classical states peaked around the expectation values for the $\u(N)$ generators and the scalar product operators:
    \be
    \la E_{ij}\ra= 2J\,\f{\la z_i|z_j\ra}{\sum_k \la z_k|z_k\ra}=\f{J}{A(z)}\la z_i | z_j\ra,
    \ee
    %We can also compute the expectation of the
    \be \label{expectJJ}
    \forall i\ne j,\quad 4\,\la \vJ_i\cdot\vJ_j\ra\,=\, \f{J^2}{A(z)^2}\vV(z_i)\cdot\vV(z_j)
    +\f{J}{2\,A(z)^2} \left(\vV(z_i)\cdot\vV(z_j)-3|\vV(z_i)|\,|\vV(z_j)|\right).
    \ee

\item The scalar product between two coherent states is easily computed:
$$
\la J,\{z_{i}\}|J,\{w_{i}\}\ra = \mathrm{det}\left(\sum_{i}|z_{i}\ra\la w_{i}|\right)^{J}
\,=\,
\left(\f12\tr \,Z\dag W\right)^J.
$$

\item They minimize the uncertainties on the $E_{ij}$ operators. The interested reader can find the various uncertainties computed in \cite{UN3}. The simplest is the $\U(N)$-invariant uncertainty:

\be
\Delta
\,\equiv\,
\sum_{ij}  \la E_{ij}E_{ji}\ra- \la E_{ij}\ra \la E_{ji}\ra
\,=\,
2J\,(J+N-2)-2J^2
\,=\,
2J\,(N-2).
\ee

%\item

\item They are related to the coherent  intertwiners discussed above:
    \be
    \f{1}{\sqrt{J!(J+1)!}}|J,\{z_i\}\ra
    \,=\,
    \sum_{\sum j_i=J}
    \frac{1}{\sqrt{(2j_{1})!\cdots (2j_{N})!}} \,||\{j_i,z_i\}\ra.
    \label{sumcoh}
    \ee

\item The coherent states $|J,\{z_i\}\ra$ at fixed $J$ provide an over-complete basis on the space $\cH_N^{(J)}$. This gives a new decomposition of the identity on that space $\id_N^{(J)} = \int [d\mu(z_i)]\,|J, \{z_{i}\}\ra\la J,\{z_{i}\}|$ where $[d\mu(z_i)]$ is a $\U(N)$-invariant measure (on $\C\mathbb{P}_{2N-1}$). For more details, the interested reader can refer to \cite{UN3}.
    %{\bf UPDATE THIS LATER}

\end{itemize}

%%%
\subsection{Relaxing the Closure Conditions}
%%%

Discussing the $\U(N)$ coherent states in the previous section, we have assumed that the spinor labels satisfy the closure condition \Ref{closure} that requires that $\sum_i \vV(z_i)=0$ or equivalently that $\sum_i |z_i\ra\la z_i| \propto \id$, or even equivalently that the two components of the spinors $z^0_i$ and $z^1_i$ are orthonormal $N$-vectors. It has been shown in \cite{UN3} how to relax this closure condition using the $\SL(2,\C)$ invariance of the coherent states. Let us review this procedure.

We consider the $\GL(2,\C)$ action acting simultaneously on all spinors $z_i$. It has a simple rescaling action on the $Z_{ij}$ matrix, which means that the $\U(N)$ coherent states also get simply rescaled:
\be
\forall \Lambda\in\GL(2,\C),\quad z_i\arr \Lambda z_i,\qquad
Z_{ij}\arr \det\Lambda\, Z_{ij},\qquad
|J,\{z_i\}\ra\arr (\det\Lambda)^J\,|J,\{z_i\}\ra\,.
\ee
Thus two coherent states labeled by spinors related through a $\GL(2,\C)$ action define the same quantum state, up to normalization. In particular, if the transformation $\Lambda$ lies in $\SL(2,\C)$ then the coherent state is exactly the same. The moot point is that $\GL(2,\C)$ transformations allow to go in and out of the closure constraint. Indeed, following \cite{UN3}, given an arbitrary set of $N$ spinors, we consider the matrix:
\be
X(z)\,\equiv\,\sum_i |z_i\ra\la z_i|.
\ee
Since $X(z)$ is obvious a positive Hermitian matrix, there exists a matrix $\Lambda\in\SL(2,\C)$ which takes its square-root, $X=\sqrt{\det X}\,\Lambda\Lambda\dag$. This matrix is unique up to $\SU(2)$ transformations. It allows to define a new set of spinors $\tz_i\equiv \Lambda^{-1}\,z_i$ which induce the same coherent state but also satisfy the closure condition:
\be
\tX=\sum_i |\tz_i\ra\la \tz_i|
\,=\,
\Lambda^{-1}\,X\,(\Lambda\dag)^{-1}
\,=\,
\sqrt{\det X}\,\id,\qquad
\det\tX=\det X.
\ee
This is the exact same $\SL(2,\C)$ action used in \cite{coh2,coh3} to take the standard coherent intertwiners in and out of the closure constraint. Let us point out that the $\SL(2,\C)$ action is simply the complexified $\SU(2)$-action still generated by the operators $J^{z,\pm}$ quadratic in the harmonic oscillators. In the $\U(N)$ framework, this simply mean that we can drop the closure condition on the spinor label, when defining $\U(N)$ coherent states and integrating over spinor labels, e.g. in the decomposition of the identity.
Moreover, the coherent states $|J,\{z_i\}\ra$ still transform covariantly under $\U(N)$ whether they satisfy the closure condition or not, and their norm is easily computed:
\be
\la J,\{z_i\}|J,\{z_i\}\ra\,=\, (\det X)^J.
\ee
Since the projectors $|z_i\ra\la z_i|$ are easily expressed in term of the classical 3-vectors $\vec{V}(z_i)$, we give similar expressions for the matrix $X$ and its determinant:
\be
X=\f12\left(
\sum_i |\vec{V}(z_i)|\,\id
+\sum_i \vec{V}(z_i)\cdot\vec{\sigma}
\right)
\quad\Rightarrow\quad
\det X
\,=\,
\f14\left(
\left(\sum_i |\vec{V}(z_i)|\right)^2 -\left|\sum_i \vec{V}(z_i)\right|^2
\right),
\ee
so that $(\det\,X)^J=A(z)^{2J}$ as before when the closure condition $\sum_i \vec{V}(z_i)=0$ is satisfied. Let us underline that $\det\,X\ge0$ can be interpreted as a measure of how far from the closure condition we are: the larger the total 3-vector $\sum_i \vec{V}(z_i)$ is, the smaller $\det\,X$ gets.

Finally, we can write a decomposition of the identity on the intertwiner space $\cH_N^{(J)}$ as an integral over $\C^{2N}$:
\be
\label{iduNcoh}
\id_{\cH_N^{(J)}}
\,=\,
D_{N,J}\,\int_{\C^{2N}}
\prod_i \f{e^{-\la z_i|z_i\ra}\,d^4z_i}{\pi}\,
\f{|J,\{z_i\}\ra\la J,\{z_i\}|}{\left(\det X(z)\right)^J}.
\ee
This is to be compared with the decomposition of the identity on $\cH_{j_1,..,j_N}$ in term of coherent intertwiners \Ref{idcohint}. To check this identity, it is enough to check that this integral commutes with the $\U(N)$-action and that its trace is equal to the dimension $D_{N,J}$ of the Hilbert space $\cH_N^{(J)}$. As explained in more details in \cite{UN3}, we can gauge-fix this integral by the $\GL(2,\C)$-action and restrict it to an integral over the Grassmanian space ${\rm Gr}_{2,N}=\C^{2N}/\GL(2,\C)=\U(N)/\U(N-2)\times \U(2)$. The $\SL(2,\C)$-action allows to gauge-fix to spinors satisfying the closure condition; then rescaling the state allows to fix the matrix $X(z)=\id$ and the total area $A(z)=1$ thus to restrict the integral to coherent states of unit norm.

%%%
\subsection{The $F$-action on Coherent Intertwiners}
%%%

%Discuss the action of the $F$ operators on coherent intertwiners and on $\U(N)$ coherent states and finally diagonalize the $F$ operators.

In order to discuss the simplicity constraints in the $\U(N)$ framework, we need the explicit action of the operators $E_{ij},F_{ij},F\dag_{ij}$ on the coherent states. Let us start by looking closer at the $F$ annihilation operators. We first compute the action of $F_{ij}$ on coherent intertwiners:
\be
F_{ij}\,||\{j_k,z_k\}\ra
\,=\,
\int_{\SU(2)} dg\,g\vartriangleright
F_{ij}\,\otimes_k\f{(z^0_ka\dag_k+z^1_kb\dag_k)^{2j_k}}{\sqrt{(2j_k)!}}\,|0\ra,
\ee
since the operator $F_{ij}$ is invariant under global $\SU(2)$ transformations and thus commutes with the action of group elements $g\in\SU(2)$. Making $F_{ij}=(a_ib_j-a_jb_i)$ commute through the creation operators, we obtain after a straightforward calculation:
\be
F_{ij}\,||\{j_k,z_k\}\ra\,=\,
Z_{ij}\,\sqrt{(2j_i)(2j_j)}\,
||\{j_i-\f12,..,j_j-\f12,j_k,z_k\}\ra,
\ee
where we remind the reader that $Z_{ij}=(z^0_iz^1_j-z^1_iz^0_j)$.
Then using the formula \Ref{sumcoh} of $\U(N)$ coherent states in term of coherent intertwiners, we easily get:
\be
F_{ij}\,|J,\{z_k\}\ra
\,=\,
F_{ij}\,\sum_{\sum j_k=J}
\frac{\sqrt{J!(J+1)!}}{\sqrt{(2j_k)!}} \,||\{j_k,z_k\}\ra
%\,=\,
%Z_{ji} \sum_{\sum j_k=J-1}
%\frac{\sqrt{J!(J+1)!}}{\sqrt{(2j_k)!}} \,||\{j_k,z_k\}\ra
\,=\,
Z_{ij}\sqrt{J(J+1)}\,|J-1,\{z_k\}\ra.
\ee
This fits with the $F$-action on $\U(N)$ coherent states derived in \cite{UN3}. Moreover we can use these relations to diagonalize the $F_{ij}$ operators. We introduce the vectors $|\beta,\{z_k\}\ra$ for $\beta\in\C$~:
\be
|\beta,\{z_k\}\ra\,\equiv\,
\sum_{J\in\N}\f{\beta^{2J}}{\sqrt{J!(J+1)!}}\,|J,\{z_k\}\ra
\quad\Rightarrow\quad
F_{ij}\,|\beta,\{z_k\}\ra\,=\,\beta^2 Z_{ij}\,|\beta,\{z_k\}\ra.
\ee
These new intertwiners $|\beta,\{z_k\}\ra$  diagonalize all $F_{ij}$ operators simultaneously. This is normal since the $F_{ij}$'s all commute with each other. We can also give other convenient expressions for these vectors in term of creation operators acting on the vacuum:
\bes
|\beta,\{z_k\}\ra
&=&
\sum_J \f{(\beta^2F\dag_Z)^J}{J!(J+1)!}\,|0\ra\\
&=&
\int dg\, g\vartriangleright
\otimes_k e^{\beta(z^0_ka\dag_k+z^1_kb\dag_k)}\,|0\ra,
\ees
which makes a clear link between these vectors and coherent states for the harmonic oscillator. Finally, we can compute the norm of these states, which is easily expressed as a Bessel function:
\be
\la\beta,\{z_k\}|\beta,\{z_k\}\ra
=\sum_J \f{(|\beta|^2)^{2J}}{J!(J+1)!} \,\la J,\{z_k\}|J,\{z_k\}\ra
=\sum_J \f{(|\beta|^2A(z))^{2J}}{J!(J+1)!}
=\f{I_1(2|\beta|^2A(z))}{|\beta|^2A(z)},
\ee
where we assumed the closure condition on the spinors so that the norm of the $\U(N)$ coherent state is given directly by $A(z)^{2J}$ (else we should in general replace $A(z)$ by the determinant $\sqrt{\det\,X(z)}$).
Here $I_1$ is the first modified Bessel function of the first kind. This clears up the action of the $F$-operators. Below, we further investigate the actions of the $E$ and $F\dag$ operators on the $\U(N)$ coherent states.

%%%
\subsection{Operator Algebra on Coherent Intertwiners}
%%%

We already have the action of the annihilation operators $F_{ij}$ on the $\U(N)$ coherent states. Now we need to complete the algebra to derive the action of the operators $F\dag_{ij}$ and $E_{ij}$. To this purpose, we use the standard action as differential operators of the creation and annihilation operators for the harmonic oscillator (see in appendix for some details):
$$
a_i \arr z^0_i,\qquad a\dag_i \arr \f{\pp}{\pp z^0_i},\qquad
b_i \arr z^1_i,\qquad b\dag_i \arr \f{\pp}{\pp z^1_i}.
$$
Using this on the definition of the operators $E$ and $F$, we guess the following action of these operators on the $\U(N)$ coherent states:
\bes
F_{ij} |J,\{z_k\}\ra&=& \sqrt{(J+1)J}Z_{ij} |J-1,\{z_k\}\ra,\\
F^\dagger_{ij} |J,\{z_k\}\ra&=&\f{1}{\sqrt{(J+2)(J+1)}} \Delta^z_{ij}|J+1, \{z_k\}\ra, \\
E_{ij}|J,\{z_k\}\ra&=&\delta^z_{ij}|J,\{z_k\}\ra,
\ees
where $Z_{ij}=(z^0_iz^1_j-z^1_iz^0_j)$ as before and where we have defined the following differential operators with respect to the spinor $z_i$:
\bes
\Delta^z_{ij}&= &\f{\pp^2}{\pp z_i^{0}\pp z_j^{1}}-\f{\pp^2}{\pp z_i^{1}\pp z_j^{0}},\\
\delta^z_{ij}&=&z_j^{0}\f{\pp}{\pp z_i^{0}}+z_j^{1}\f{\pp}{\pp z_i^{1}}.
\ees
The $J$-factors in front of the actions of $F$ and $F\dag$ come from the normalization factor $\sqrt{J!(J+1)!}$ of the coherent states.

The multiplication action of $F$ on the $\U(N)$ coherent states can be derived by using the commutation relation between the creation and annihilation operators:
\bes \label{Fact_1}
&&\left[ F_{ij}, F_Z^\dagger \right]= \mathcal{E}^Z_{ij} +2Z_{ij},\\
&&\left[\mathcal{E}^Z_{ij}, F^\dagger_Z \right]=2Z_{ij}F_Z^\dagger.
\ees
where we have defined the auxiliary operator $\mathcal{E}^Z_{ij}= \sum_m \left(Z_{im}E_{mj}-Z_{jm}E_{mi} \right)$. To show the second commutator, we have used that the antisymmetric matrix $Z$  satisfies that $Z_{ik} Z_{jl}-Z_{il} Z_{jk}= Z_{ij} Z_{kl}$. This allows the straightforward calculation:
\bes \label{Fact_2}
F_{ij}\left(F_Z^\dagger \right)^J|0\ra
&=&
\sum_{k=0}^{J-1} \left(F_Z^\dagger \right)^{J-1-k} \mathcal{E}^Z_{ij} \left( F^\dagger_Z\right)^k|0\ra +2J Z_{ij} \left(F_Z^\dagger \right)^{J-1}|0\ra, \nn\\
&=&
\left(\sum_{k=0}^{J-1} 2k +2J\right) Z_{ij} \left(F_Z^\dagger \right)^{J-1}|0\ra, \nn\\
&=&
J(J+1) Z_{ij}\left(F_Z^\dagger \right)^{J-1} |0\ra,
\ees
which gives the expected result. Moreover, we recover the same action for the $F_{ij}$ operators that we had already computed in the previous section using that the $\U(N)$ coherent states are superpositions of coherent intertwiners.

As for the $F\dag$-action, it is straightforward to compute the action of the differential operator $\Delta^z_{ij}$ on the coherent state taking into account that:
\be
\Delta^z_{ij}\,(Z_{kl})=2(\delta_{ik}\delta_{jl}-\delta_{il}\delta_{jk}),\qquad
\Delta^z_{ij}\,(F\dag_Z)^J
\,=\,
\Delta^z_{ij}\,\left(\f12\sum_{kl}Z_{kl}F\dag_{kl}\right)^J
\,=\,
J(J+1)F\dag_{ij}\,(F\dag_Z)^{J-1}.
\ee
This leads to the following action for the creation operator $F\dag_{ij}$
\be
\Delta^z_{ij} |J+1, \{z_k\} \ra= \sqrt{(J+1)(J+2)}F\dag_{ij}\,|J, \{z_k\}\ra,
\ee
since $F\dag_{ij}$ commutes with $F\dag_Z$ because they only involve oscillator creation operators $a\dag$ and $b\dag$.
At this stage, we can also check that the differential $F\dag$-action is indeed the adjoint of the multiplicative  $F$-action on the $\U(N)$ coherent state. That is straightforward to show. First, considering the matrix element $\la K, \{w_k\}| F_{ij}|J,\{z_k\}\ra$, it doesn't vanish iff $K=(J-1)$. Then, on the one hand,  we can compute:
\be
\la J-1, \{z_k\}| F_{ij}|J,\{w_k\}\ra
=\sqrt{J(J+1)}\, W_{ij} \,\la J-1,\{z_k\}|J-1, \{w_k\}\ra
= \sqrt{J(J+1)}\, W_{ij} \,\left(\f12\tr\,Z\dag W\right)^{J-1}.
\ee
On the other hand, we have:
\be
\la J,\{w_k\} |F\dag_{ij}|J-1, \{z_k\}\ra
=\f{1}{\sqrt{J(J+1)}}\,\Delta^z_{ij}\,\la J,\{w_k\} |F\dag_{ij}|J, \{z_k\}\ra
=\f{1}{\sqrt{J(J+1)}}\,\Delta^z_{ij}\,\left(\f12\tr\,W\dag Z\right)^{J}.
\ee
To evaluate this expression, we calculate explicitly the action of the differential operator on the $J$-power of the trace:
\be
\Delta^z_{ij}\,\left(\tr\,W\dag Z\right)^{J}
\,=\,
2J(J+1)\,\overline{W}_{ij}\left(\tr\,W\dag Z\right)^{J-1}.
\ee
This allows to conclude that we have as expected:
\be
\overline{\la J-1, \{z_k\}| F_{ij}|J,\{w_k\}\ra}=\la J,\{w_k\} |F\dag_{ij}|J-1, \{z_k\}\ra.
\ee

Finally, let us now compute the action of the $E$-operators on the U(N) coherent states. First we compute the commutator $\left[ E_{ij}, F_Z^\dagger\right] = \sum_{m} Z_{jm} F^\dagger_{im}$, which easily gets generalized to arbitrary power of the creation operator:
\be
[E_{ij},(F_Z^\dagger)^J]
\,=\,
%(F_Z^\dagger\right)^J\,E_{ij}+
\sum_{k=0}^{J-1} (F_Z^\dagger)^{J-1-k}
\left[ E_{ij}, F_Z^\dagger\right]
(F_Z^\dagger)^k
\,=\,
J\,\left(\sum_{m} Z_{jm} F^\dagger_{im}\right)\,
(F_Z^\dagger)^{J-1},
\ee
since all $F\dag$ commute with each other. This proves directly that the $E$-action on $\U(N)$ coherent states is simply related to the $F\dag$-action:
\be
E_{ij} (F_Z^\dagger)^{J}\,|0\ra
\,=\,
J\,\left(\sum_m Z_{jm} F\dag_{im}\right)\, (F_Z^\dagger)^{J-1}\,|0\ra.
%E_{ij} |J, \{z_k\}\ra= \sqrt{\f{J}{J+1}}\,\left(\sum_m Z_{jm} F\dag_{im}\right)\, |J-1,\{z_k\}\ra
\ee
Then we can easily compute the action of the differential operator:
\bes
\delta^z_{ij} \left(F_Z^\dagger \right)^J&=& \left(z_j^{0}\f{\pp}{\pp z_i^{0}}+z_j^{1}\f{\pp}{\pp z_i^{1}}\right) \left(\f12\sum_{kl} Z_{kl}F\dag_{kl} \right)^J \nn \\
&=& J\sum_{m}\left(Z_{jm} F^\dagger_{im}\right) \left(F_Z^\dagger \right)^{J-1}
\ees
This allows us to deduce the actions of the $E$-operators and of the differential operators $\delta^z_{ij}$ match on the $\U(N)$ coherent states:
\be
E_{ij}|J,\{z_k\}\ra=\delta^z_{ij}|J,\{z_k\}\ra.
\ee
It is possible to check directly that these differential operators actually generate the correct $\U(N)$ action on the spinors. Let us for instance consider the infinitesimal unitary transformation $u=\exp(\eps(E_{ij}-E_{ji}))$ where $i,j$ are arbitrary but fixed indices. It acts at first order on the spinors as:
$$
(u\,z)_k\sim z_k + \eps\delta_{ik}z_j - \eps\delta_{jk}z_i.
$$
It is very easy to check that this fits with the action of the previous differential operator:
$$
\eps(\delta^z_{ij}-\delta^z_{ji})\,z_k
\,=\,
 i \eps\delta_{ik}z_j - \eps\delta_{jk}z_i
\,\sim\,
(u\,z)_k- z_k.
$$
Following the same steps with the unitary transformations $u=\exp(i\eps(E_{ij}+E_{ji}))$ allows to prove completely that the differential operators $\delta^z_{ij}$ generate as expected the $\U(N)$ action on our coherent states.

Finally, it is also possible to check that the commutation relation between the differential operators corresponding to $E$, $F$ and $F\dag$ satisfy the correct commutation relations (see in appendix).

%%%
\subsection{The $\U(N)$ setting for $\spin(4)$ intertwiner and Diagonal Simplicity}
%\subsection{Solving Weakly the Simplicity Constraints}
%%%

We have reviewed the whole $\U(N)$ formalism for $\SU(2)$ intertwiners and we gave the explicit action of the operators $E_{ij},F_{ij},F\dag_{ij}$ on the $\U(N)$ coherent states. Now, we come back to the initial problem and to $\Spin(4)$ intertwiner. Since the Lie algebra $\spin(4)=\su_L(2)\oplus\su_R(2)$ simply splits into two copies of the $\su(2)$ algebra, it is straightforward to adapt the whole $\U(N)$ framework to $\spin(4)$. We double all the operators, introduce harmonic oscillators $a_i^L,b_i^L$ and $a_i^R,b_i^R$ and build two sets of $\u(N)$ operators $E^L_{ij},F^L_{ij},F^{L\dagger}_{ij}$ and $E^R_{ij},F^R_{ij},F^{R\dagger}_{ij}$.
These two $\u(N)$ sectors don't speak to each other and are a priori decoupled. It is the simplicity constraints that will couple them.

Let us start with the diagonal simplicity constraints. They impose that the $\spin(4)$ living on the legs of the intertwiners are simple. This means that the spins in the left and right sectors  match: $j_i^L=j_i^R$. This translates into very simple constraints in the $\u(N)$ framework:
\be
\cC_i\,\equiv\,
E^L_{i}-E^R_{i}.
\ee
This diagonal simplicity  definitively couples the two sectors. This constraint is actually the same than the matching conditions for spin networks on the 2-vertex graph and the whole construction will be very similar \cite{2vertex}. Every (constraint) operator that we will now introduce to solve the simplicity constraints will have to commute (at least weakly) with these diagonal simplicity constraints $\cC_i$.

Now moving to the crossed simplicity constraints, they refer to couples of legs of the intertwiners and to their scalar product. They amount to impose strongly, weakly or semi-classically, the equality between the scalar products of the left and right sectors, $\vJ_i^L\cdot\vJ_j^L=\vJ_i^R\cdot\vJ_j^R$. Dropping the $L/R$ index, we remind the expression for the scalar product operator in term of $\u(N)$ operators for $i\ne j$:
\bes \label{scalarProdOp}
%\forall i\ne j,\,\,
\vJ_i\cdot \vJ_j
&=&
\f12E_{ij} E_{ji} -\f14E_iE_j-\f12 E_i,\\
&=&
\f12E_{ji} E_{ij} -\f14E_iE_j-\f12 E_j,\nn\\
&=& -\f12 F_{ij}^\dagger F_{ij} +\f14 E_iE_j,\nn\\
&=& -\f12 F_{ij}F_{ij}^\dagger +\f14 (E_i+2)(E_j+2).\nn
\ees
This expression clearly suggests two things. First, we could replace the $\vJ_i^L\cdot\vJ_j^L=\vJ_i^R\cdot\vJ_j^R$ constraints by some constraints of the type $E^L=E^R$ or $F^L=F^R$. We will explore these various possibilities below. Second, we then expect that the equality $\vJ_i^L\cdot\vJ_j^L=\vJ_i^R\cdot\vJ_j^R$ will only hold semi-classically at first order and will usually have corrections linear in the $j_i$'s (terms in $E_i$ and $E_j$).

 %%%%%%%%%%%%%%%%%%%%%%%%%%N'ai rien modifi dans tout ce qui prcde par rapport  l'ancienne version
%%%%%%%%%%%%%%%%%%%%%%%%%%%%%%%%%%%%%%%%%%%%%%%%%%%%%%%
\section{The New $\U(N)$ Simplicity Constraints}
%%%%%%%%%

%%%
\subsection{The Closed Algebra of Simplicity Constraints}
%%%

%\begin{itemize}
%
%\item $\U(N)$ operators explicitly introduced to close the algebra of scalar product operators, should be relevant here where the main issue is that the alg of simplicity constraints does not close!
%
%\item discuss the two copies of the $E,F$ alg, and the two sets of coupled $\U(N)$ operators
%
%\item Thus introduce the $E^L-(E^R)\dag$ constraints, closed alg, discuss diagonal ops and off-diagonal ops
%
%\item Discuss resulting relations on expectation values
%
%\item Solve new $\U(N)$ Simplicity Constraints by group averaging?
%
%\end{itemize}

One big issue about the standard crossed simplicity constraints $\vJ_i^L\cdot\vJ_j^L-\vJ_i^R\cdot\vJ_j^R=0$ for all couples of legs $i\ne j$ is that their algebra doesn't close. The $\U(N)$ framework was precisely introduced to close the algebra of scalar product operators and provide an alternative algebra for invariant observables on the space of intertwiners. Indeed, considering the operators $E_{ij}$ instead of $\vJ_i\cdot \vJ_j$ allowed to have a closed algebra of invariant observables and to build coherent intertwiner states \'a la Perelomov that transforms nicely under the operators of that algebra.
This naturally suggests to replace the simplicity constraints $\vJ_i^L\cdot\vJ_j^L-\vJ_i^R\cdot\vJ_j^R=0$ by a simpler constraint expressed in term of the operators $E_{ij}^{L,R}$.
We propose to consider a new set of constraints, that we name the $\u(N)$ simplicity constraints:
\be
\cC_{ij}\,\equiv\,
E^L_{ij}-E^R_{ji}=E^L_{ij}-(E^R_{ij})\dag,
\qquad\qquad
\cC\dag_{ij}=\cC_{ji}.
\ee

The two important facts about these new proposed contraints are:
\begin{itemize}

\item They naturally include the diagonal simplicity constraints:
$$
\cC_{ii}=\cC_i=E^L_{i}-E^R_{i}.
$$

\item They form a closed $\u(N)$ algebra:
\be
[\cC_{ij},\cC_{kl}]
\,=\,
\delta_{jk}\cC_{il}-\delta_{il}\cC_{kj}.
\ee

\end{itemize}

Moreover, let us $\cH_\cC$ be the Hilbert space  of states satisfying these $\u(N)$ simplicity constraints:
\be
\cH_\cC
\,\equiv\,
\{|\psi\ra\,\textrm{ such that }\, \cC_{ij}\,|\psi\ra=0,\,\forall i,j\}.
\ee
Then this solves weakly the crossed simplicity constraints at leading order (i.e for large spins). Indeed, for all solution states $\phi,\psi\in\cH_\cC$, we have for $i\ne j$:
\bes
\la\phi|\vJ_i^L\cdot\vJ_j^L|\psi\ra
&=&
\la\phi|\f12E_{ij}^L E_{ji}^L -\f14E_i^LE_j^L-\f12 E_i^L|\psi\ra
\,=\,
\la\phi|\f12E_{ji}^R E_{ij}^R -\f14E_i^RE_j^R-\f12 E_i^R|\psi\ra,\nn\\
%\,=\,
%\la\phi|\f12E_{ji}^R E_{ij}^R -\f14E_i^RE_j^R-\f12 E_j^R+\f12(E_j^R-E_i^R)|\psi\ra
&=&
\la\phi|\vJ_i^R\cdot\vJ_j^R|\psi\ra\,+\,\la\phi|\f12(E_j^R-E_i^R)|\psi\ra.
\ees
Therefore, the crossed simplicity constraints are solved approximatively at first order. Indeed the expectation values $\la\vJ_i^L\cdot\vJ_j^L\ra$ are of order $\cO(j^2)$ while the correction term is of order $(j_j-j_i)\sim\cO(j)$.
This is not a very big obstacle since we only expect the simplicity constraints to be satisfied semi-classically in the large spin regime. Let us still point out that the diagonal simplicity constraints are still strongly and exactly enforced on all invariant states in $\cH_\cC$.

\medskip

As we said above, the operators $\cC_{ij}$ generate a $\u(N)$ Lie algebra: they actually generate the $\U(N)$ action $(u,\bar{u})$ on the coupled $L,R$ system such that the $\U(N)$ transformation acting on the right sector is the complex conjugate  of the transformation acting on the left sector.
Indeed, a finite transformation generated by the constraints $\cC_{ij}$ will read, for a anti-Hermitian matrix $\alpha_{ij}=-\bar{\alpha}_{ji}$:
$$
U
\,\equiv\,
e^{\sum_{ij}\alpha_{ij}\cC_{ij}}
\,=\,
e^{\sum_{ij}\alpha_{ij}E_{ij}^L}\,e^{-\sum_{ij}\alpha_{ij}E_{ji}^R}
\,=\,
e^{\sum_{ij}\alpha_{ij}E_{ij}^L}\,{e^{\sum_{ij}\overline{\alpha_{ij}}E_{ij}^R}}.
$$
%since the $\u(N)$ operators are Hermitian, $E_{ji}^R=E_{ij}^R{}\dag$.
%
%Let consider an arbitrary group element $U= e^{\sum_{ij}\alpha_{ij}\cC_{ij}} \in \U(N)$ where $\alpha$ is an anti-hermitian matrix and $|\psi \ra \in \cH_\cC$, then $U|\psi\ra= (1 +\sum_{p>0}\f{(\sum_{ij} \alpha_{ij} \cC_{ij})^p}{p!} ) |\psi\ra= |\psi\ra$.
%
Thus, states which are solution to the $\cC_{ij}$-constraints are $\U(N)$-invariant and the Hilbert space $\cH_\cC$ can be obtained by performing a $\U(N)$ group averaging on the space of intertwiners $\bigoplus_J \cH_N^{(J),L}\otimes \cH_N^{(J),R}$.
An over-complete basis of solutions can be obtained by group averaging the $\U(N)$ coherent states $|J,\{z^L_k \} \ra \otimes |J, \{ z^R_k \} \ra$.
However, we can give a more precise description of the $\U(N)$-invariant space. Indeed, since the spaces $\cH_N^{(J)}$ are irreducible $\U(N)$-representations, the Schur's lemma implies that there exists a unique invariant vector in the tensor product $\cH_N^{(J), L}\otimes \cH_N^{(J),R}$. Calling $|J\ra$  this unique state solution to the $\u(N)$-constraints for every total spin $J$, we have a complete basis of our solution space:
\be
\cH_\cC =\bigoplus_J \C\,|J\ra.
\ee
This construction is exactly the same than the definition of isotropic states in the 2-vertex loop quantum gravity model constructed in \cite{2vertex} using the $\U(N)$ formalism. Thus following that approach, we won't perform the $\U(N)$-group averaging to construct our $\U(N)$-invariant states but we will use using the following symmetric operator~:
\be
f^\dagger \equiv \sum_{kl} F_{kl}^{L \dagger} F_{kl}^{R \dagger}.
\ee
Indeed, this operator commute with all generators $\cC_{ij}$:
\be
\left[\cC_{ij}, f^{ \dagger}  \right] = \sum_{kl} \left(\left[ E_{ij}^L, F^{L \dagger}_{kl}\right] F^{R \dagger}_{kl} - F^{L \dagger}_{kl} \left[ E^R_{ji}, F^{R \dagger}_{kl} \right] \right) =0,
\ee
therefore,  the operator $f^{ \dagger}$ is $\U(N)$-invariant.
%
%We can also check this directly by applying a finite unitary transformation to $f\dag$~:
%$$
%U\vartriangleright f\dag
%\,=\,
%$$
%
Since the vacuum state is also $\U(N)$-invariant, we can define the invariant states by applying this creation operator $f\dag$ to the vacuum state $|0\ra$~:
\be
|J\ra \equiv \left(f^\dagger \right)^J |0\ra
\ee
is obviously $\U(N)$-invariant. We also check that $|J\ra \in \cH_N^{(J),L}\otimes \cH_N^{(J),R}$.
Indeed, a straightforward calculation of the action of the total spin operator  $E \equiv \sum_i E^{L}_i=\sum_i E^{R}_i$ (the left and right total spin operators are obviously equal on the invariant space $\cH_\cC$) gives~:
\be
E |J\ra=2J |J\ra.
\ee
Finally, following the computations done in \cite{2vertex}, we also give the norm of these invariant vectors:
\be
\la J | J \ra= 2^{2J} J! (J+1)! \f{(N+J-1)!(N+J-2)!}{(N-1)! (N-2)!}=2^{2J}(J!(J+1)!)^2 D_{N,J}
\ee
where $D_{N,J}$ is the dimension of the intertwiner space $\cH_N^{(J)}$ given by (\ref{dimNJ}). The details of this calculation can be found in the appendix.

The fact that we get a single state for each total spin means that we are imposing constraints which are too strong. In the next parts, we will try too impose less constraints using the $E$ operators then different constraints in terms of the $F$ and $F^\dagger$ operators in order to get a bigger set of solutions to the simplicity constraints. Finally, we will see in the last section how we can use the $\U(N)$ coherent states in order to solve weakly all the simplicity constraints.
%So they are explicitly first class constraints and can be solved through group averaging as we show in the next section. More precisely, the operators $\cC_{ij}$ generates a $\U(N)$ action $(u,\bar{u})$ on the coupled $L,R$ system such that the $\U(N)$ transformation acting on the right sector is the complex conjugate  of the transformation acting on the left sectors.

%There exists another set of coupled $\U(N)$ transformations $(u,u)$ which acts the same way on both sectors and that are generated by the symmetric operators $e_{ij}=E^L_{ij}+E^R_{ij}$. It is obvious to check that these operators $e_{ij}$ form a closed $\u(N)$ algebra.
%
%If we were using these operators as new simplicity constraints $e_{ij}=0$, we would solve weakly the  exact simplicity constraints without correction terms.
%
%However, we face two issues:
%\begin{itemize}

%\item These constraints are not compatible with the diagonal simplicity constraints $\cC_k$:
%\be
%[E^L_{ij}+E^R_{ij},\cC_{k}]= \delta_{jk}(E^L_{il}-E^R_{il})-\delta_{il}(E^L_{kj}-E^R_{kj}).
%\ee
%Therefore, we lose the interesting features that the set of all constraints  form a closed algebra.

%\item If we consider the diagonal part of the $e_{ij}$ algebra, we have the operators $e_i=E^L_{i}+E^R_{i}$. If we impose these constraints, we have a unique solution, which is the vacuum state $|0\ra$, since the operators $E^{L,R}_{i}$ are strictly positive operators (and give the spin $j_i\in\N/2$).

%\end{itemize}
%For these reasons, we discard the possibility of considering this set of operators $e_{ij}$ as new simplicity constraints.

%%%
\subsection{Highest weight vectors for the $\u(N)$-simplicity constraints}
%%%

As we said in the previous section, it seems that the $\u(N)$-simplicity constraints are too strong. Following the idea that we might have imposed too many constraints, we propose  a new restricted set of $\u(N)$ constraints and consider only the raising operators of our $\u(N)$ algebra. This is also consistent with the line of thoughts that such a procedure usually leads to the construction of proper coherent states with the expected semi-classical properties.
Thus we try with the following new set of constraints:
\be \label{Erelax}
\{ \cC_{i<j}\equiv \cC_{ij} \textrm{ for } i< j \textrm{ and } \cC^\sigma_i=\cC_i-\sigma_i  \}
\ee
where we have chosen to require that only the raising operators\footnotemark{ }vanish
$\cC_{ij}\,|\psi\ra\,=0$ for $i<j$. We have also relaxed the diagonal simplicity constraints:  $\cC_i\,|\psi\ra\,=\sigma_i\,|\psi\ra$ where the parameters $\sigma_i\in\Z/2$ are arbitrary but fixed. In general, we will require that  $|\sigma_i|<< j_i$, so that the diagonal simplicity constraint are still satisfied at leading order.
\footnotetext{
This new set of  constraints \ref{Erelax} still forms a closed algebra. Indeed, let  be $i\leq j$ and $k\leq l$ then:
\be
\left[\cC_{ij}, \cC_{kl} \right]= \delta_{jk} \cC_{il} -\delta_{il} \cC_{kj}
\ee
where $i \leq j, \, k \leq l$ and $j=k$ imply $i \leq l$ or $k \leq l, \, i \leq j$ and $i=l$ imply $k\leq l$. Therefore, $\cC_{il}$ and $\cC_{kj}$ are also raising operators.}

Even we do not impose the full $\u(N)$ simplicity constraints, the cross simplicity constraints are still weakly satisfied. Indeed, let us define the Hilbert space $\cH_{\cC^<_\sigma}$ of states which satisfy the restricted set of constraints (\ref{Erelax}).  For all states $\phi, \, \psi \in \, \cH_{\cC^<_\sigma}$, we have:
\be
\forall i<j, \qquad  \la\phi|\vJ_i^L\cdot\vJ_j^L|\psi\ra
\,=\,
\la\phi|\vJ_i^R\cdot\vJ_j^R|\psi\ra + \la\phi|\f12 (E_i^R-E_j^R)-\f14 (\sigma_i E_j^R +\sigma_j E_i^R + \sigma_i \sigma_j)-\f12 \sigma_j |\psi\ra
\ee
%\bes
%\textrm{if } i<j, \quad \ \la\phi|\vJ_i^L\cdot\vJ_j^L|\psi\ra
%&=&
%\la\phi|\vJ_i^R\cdot\vJ_j^R|\psi\ra + \la\phi|\f12 (E_i^R-E_j^R)-\f14 (\sigma_i E_j^R +\sigma_j E_i^R + \sigma_i \sigma_j)-\f12 \sigma_j |\psi\ra  \nn \\
%\textrm{if } j<i, \quad \ \la\phi|\vJ_i^L\cdot\vJ_j^L|\psi\ra
%&=&
%\la\phi|\vJ_i^R\cdot\vJ_j^R|\psi\ra + \la\phi|\f12 (E_j^R-E_i^R)-\f14 (\sigma_i E_j^R +\sigma_j E_i^R + \sigma_i \sigma_j)-\f12 \sigma_i |\psi\ra
%\ees
Therefore, the weak cross simplicity constraints are still satisfied approximatively at leading order: the matrix elements $\la \phi |\vJ_i^R \cdot \vJ_j^R | \psi \ra$ are of order $\cO(j^2)$ while  the correction terms are of order $\cO(j)$.

The meaning of the Hilbert space $\cH_{\cC^<_\sigma}$ is straightforward in term of the theory of representations of the $\u(N)$ Lie algebra: it is the space of highest weight vectors. More precisely, let us consider the full space of $\spin(4)$ intertwiners defined as the tensor product of the uncoupled intertwiner spaces for $\su(2)_L$ and $\su(2)_R$. It is given by the direct sum over possible total area labels $J_L,J_R$ of the corresponding irreducible $\u(N)$ representations:
\be
\cH^{\spin(4)}_N
\,=\,
\bigoplus_{J_L,J_R}
\cH_N^{J_L}\otimes \cH_N^{J_R}.
\ee
Now our constraint algebra generates the diagonal $\u(N)$ action  which acts simultaneously on both the left and right sectors. Then we decompose the tensor products $\cH_N^{J_L}\otimes \cH_N^{J_R}$ into irreducible representations of this diagonal $\U(N)$ action and the vectors that are annihilated by the raising operators $\cC_{i<j}$ are the highest weight vectors  of these irreducible representations. The parameters $\sigma_i$ are the eigenvalues of the diagonal $\u(N)$ generators, they are the values of the highest weight and select the relevant representations.

For instance, the most natural case, $\sigma_i =0,\,\forall i$ , corresponds to $\U(N)$-invariant representations and we recover the space $\cH_\cC$ considered in the previous section. Then for a generic choice of $\sigma_i$, we do not necessarily require that $J_L=J_R$ as before, but this condition is slightly shifted to $J_L=J_R +\sum_i\sigma_i$.
The next step would be to decompose the product tensor of the two $\U(N)$ representations $\cH_N^{J_L}\otimes \cH_N^{J_R}$ into $\U(N)$ irreducible representations and then to extract the highest weight vector of this decomposition which correspond to our choice of $\sigma_i$'s. This can be done using the Gelfand-Zetlin basis and the Gelfand patterns \cite{tensorUN}. We do not investigate further in this direction in this present work and we postpone such an analysis to future work.

%%%%%%%%%
%\section{Holomorphic Constraints?}
%%%%%%%%%

%%%
\subsection{Using $F^L-F^R$ Constraints} \label{Fconstraints}
%%%

%\begin{itemize}
%
%\item The most natural if looking at the crossed simplicity constraints, easy to solve once the $F$-ops are diagonalized!
%
%\item Problem with diagonal simplicity constraints
%
%\end{itemize}

Another possibility to identify new simplicity constraints within the $\U(N)$ framework is to use the $F_{ij}$-operators instead of the $E_{ij}$ operators. Moreover, introducing simplicity constraints defined in terms of the $F_{ij}^{L,R}$ would be more in the spirit of the Gupta-Bleuler procedure since the $F$'s are indeed the annihilation operators.
Following this intuition, we define $F$-constraints:
\be
f_{ij}
\,\equiv\,
F_{ij}^{L}-F_{ij}^{R}.
\ee
First, these constraints all commute with each other, $[f_{ij},f_{kl}]=0$.
Moreover, these constraints are straightforward  to solve since we know how to diagonalize explicitly and simultaneously the operators $F_{ij}$ using the superposition of coherent states $|\beta,\{z_i\}\ra$.

Furthermore, solving these constraints seem to allow to solve weakly  the exact original quadratic simplicity constraints (without correction terms). Indeed, for all states $\phi,\psi$ in the kernel of $f_{ij}$ for all $i\ne j$, we have
\bes
\la\phi|\vJ_i^L\cdot \vJ_j^L|\psi\ra
&=&
\la\phi|-\f12 F^L_{ij}{}^\dagger F^L_{ij} +\f14 E^L_iE^L_j|\psi\ra \nn\\
%\,=\,
%\la\phi|-\f12 F^R_{ij}{}^\dagger F^R_{ij} +\f14 E^L_iE^L_j|\psi\ra
&=&
\la\phi|\vJ_i^R\cdot \vJ_j^R |\psi\ra
\,+\,
\la\phi|\f14 (E^L_iE^L_j-E^R_iE^R_j)|\psi\ra.
\ees
If we also assume that the diagonal simplicity constraints hold, i.e that the operators $E^L_i-E^R_i$ vanish on both states $\psi,\phi$, then the second term vanishes and it all works out. Unfortunately, the $F$-constraints do not form a closed algebra with the diagonal constraints $\cC_i$:
\be
[\cC_i,f_{kl}]
\,=\,
[E^L_i, F^L_{kl}]+[E^R_i, F^R_{kl}]
\,=\,
\delta_{il}(F^L_{ik}+F^R_{ik})-\delta_{ik}(F^L_{il}+F^R_{il}).
\ee
Thus, if we require both $\cC_i=0$ and $F_{kl}^{L}-F_{kl}^{R}=0$, then we automatically also require $F_{kl}^{L}+F_{kl}^{R}=0$, which means that we are actually imposing the much stronger constraints $F_{kl}^{L}=F_{kl}^{R}=0$. These constraints are obviously only satisfied by the vacuum state $|0\ra$. Thus the $f_{kl}$ constraints are not consistent with the diagonal simplicity constraints. However, we will see in the last section that if we drop the requirement of imposing strongly the diagonal simplicity constraints then these $f$ constraints appear to be the right constraints to consider: they allow to impose all the (diagonal and crossed) simplicity constraints weakly.

%%%
\subsection{Using $F^L-(F^R)\dag$ Constraints}
%%%

%\begin{itemize}
%
%\item Introduce new set of constraints, check alg with diagonal simplicity constraints
%
%\item Covariance under $\U(N)$ action!
%
%\item Discuss expectation values
%
%\item Solve them??
%
%\end{itemize}

We now consider  ``holomorphic" constraints defined in terms of the $F_{ij}$ and $F_{ij}^\dagger$ operators by:
\be
\cF_{ij}\,\equiv\,
F_{ij}^{L}-F_{ij}^{R}{}\dag,\qquad
\cF_{ij}=-\cF_{ji}.
\ee

These new operators commute with each other:
\be
[\cF_{ij},\cF_{kl}]=0
\ee
and the commutator of these new constraints with the $\u(N)$ generators $\cC_{ij}$ is now given by:
\bes
[\cC_{ij},\cF_{kl}]
&=& [E_{ij}^{L}-E_{ji}^{R}\,,\,F_{kl}^{L}-F_{kl}^{R}{}\dag] \nn\\
&=& \delta_{il}\cF_{jk}-\delta_{ik}\cF_{jl}.
\ees
This shows two things. First, if we take $i=j$, the $\u(N)$ generators are the diagonal simplicity constraints. This means that the holomorphic constraints are compatible with the diagonal simplicity constraints and together they form a closed Lie algebra: we can impose $\cC_{i}=0$ on the space of solutions to $\cF=0$ without obvious obstacle. Second, let us call $\cH_\cF$ the Hilbert space of states $\psi$ satisfying $\cF_{ij}\,|\psi\ra=0$ for all indices $i,j$. Then the previous commutator also means that there is a natural $\U(N)$ action on this solution space $\cH_\cF$ generated by the operators $\cC_{ij}$. In particular, once we identify a single solution to the holomorphic constraints $\cF$ then this induces a whole family of solutions obtained by acting with $\U(N)$ transformations on that initial solution.

We introduce the Hilbert space $\mathcal{H}_{\cF}^0$ of states satisfying the holomorphic constraints and the diagonal simplicity constraints $\cC_i$. Then, for all solution states $\psi$, $\phi \, \in \cH_{\cF}^0$, the expectation values of the left and right scalar product operators are equal up to a correction of order $\cO(j)$:
\bes
\la\phi|\vJ_i^L\cdot \vJ_j^L|\psi\ra
&=&
\la\phi|-\f12 F^L_{ij}{}^\dagger F^L_{ij} +\f14 E^L_iE^L_j|\psi\ra \nn\\
&=&
\la\phi|-\f12 F^R_{ij} F^R_{ij}{}^\dagger +\f14 E^R_iE^R_j|\psi\ra \nn\\
&=&
\la\phi|\vJ_i^R\cdot \vJ_j^R |\psi\ra
\,-\,
\la\phi| 1+\f12(E_i+E_j) |\psi\ra.
\ees
To identify solution states in $\mathcal{H}_{\cF}^0$, we start by the simplest case, which is to construct $\U(N)$-invariant states solution of this new set of constraints.  We recall that while the $E_{ij}$-operators leave invariant the total sum of spins $E^{L,R}$, the $F_{ij}^{L,R}$ operators decrease by $-1$ respectively the left and right total areas $E^{L,R}$ and the $F_{ij}^{L,R}{}^\dagger$ operators increase them by $+1$. That is why we consider a linear combination of $\U(N)$-invariant states  for different areas $J$; we use the $U(N)$-invariant basis $|J\ra$. It is straightforward to compute that \footnote{The computation is similar to the computation of the multiplication action of $F$ on the $\U(N)$ coherent states, done from (\ref{Fact_1}) to (\ref{Fact_2}), replacing $Z_{kl}$ by $2 F_{kl}^{R \dagger}$: $F^{R\dagger }_{kl}$ is also antisymmetric in $k \leftrightarrow l$ and satisfies the Pl\"ucker relation ($F^{R\dagger }_{ik}F^{R\dagger }_{jl}-F^{R\dagger }_{il}F^{R\dagger }_{jk}=F^{R\dagger }_{ij}F^{R\dagger }_{kl}$). Therefore, we just recall the main steps:
$$
%\be
[F_{ij}^L, f^\dagger] = 2\underbrace{\sum_k F_{ik}^{R \dagger} E_{kj}-F_{jk}^{R \dagger} E_{ki}}_{= \mathcal{E}_{ij}^{L,F^{R\dagger}}} + 4 F_{ij}^{R \dagger}, \qquad\qquad
%\ee
%\be
[\mathcal{E}_{ij}^{L,F^{R\dagger}}, f^\dagger] =4F^{R \dagger}_{ij} f^\dagger,
%\ee
$$
therefore we get:
$$
%\be
F^{L}_{ij} |J \ra=2J(J+1) F_{ij}^{R \dagger} |J-1 \ra.
%\ee
$$
}:
\be
F^{L}_{ij} |J \ra=2J(J+1) F_{ij}^{R \dagger} |J-1 \ra.
\ee
Then if we define the states:
\be \label{alpha}
|\alpha\ra= \sum_J \f{\alpha^J}{2^J J! (J+1)!} |J\ra= \sum_J \f{\alpha^J}{2^J J! (J+1)!} (f^\dagger)^J|0\ra \quad \textrm{ with } \alpha \in \C
\ee
they satisfy:
\be
F_{ij}^L |\alpha\ra =\alpha \, F_{ij}^{R \dagger} |\alpha\ra \quad \forall \, i , j\,.
%\quad \textrm{ i.e. } \cF_{ij} |\alpha \ra =0 \quad \forall \, i,j.
\ee
Thus for $\alpha=1$, they are solution of the $\cF$-constraints: $\cF_{ij} |\alpha =1\ra =(F_{ij}^L - F_{ij}^{R \dagger})\, |\alpha=1\ra=0$. Let us notice that these new states $|\alpha\ra$ for the coupled $L/R$ system are very similar to the coherent states $|\beta,\{z_k\}\ra$ diagonalizing the $F_{ij}$ operators acting on a single (left or right) sector. It's actually the exact same expression if we replace the spinor parameters $Z_{ij}$ by the creation operators $F\dag_{ij}$ of the other sector: instead of imposing by hand the values of the expectation values using the spinor labels, the behavior of the left sector is entirely dictated by the right sector, and vice-versa. As underlined in \cite{2vertex} in the context of loop quantum gravity on the 2-vertex graph, these states $|J\ra$ and $|\alpha\ra$ maximally entangle the left and right sectors.

Therefore, we have determined the unique $\U(N)$-invariant state solution to the $\cF$ constraints.
The natural question is whether there exist  other solutions to these $\cF$-constraints, which would necessarily be non-$\U(N)$-invariant. At this point, we have not been able to identify such solutions and we would like to conjecture that they do not exist. We however postpone the precise analysis of such conjecture to future investigation. Nevertheless, we would like to point out that a promising line of tackling this issue would be to work in the coherent intertwiner basis and use the expression of the operators $E,F,F\dag$ as differential operators on the spinor labels.

\subsection{Including the Immirzi Parameter?}\label{immparam?}
%%%%%%%%%

%\begin{itemize}
%\item Imm param as a scale factor between left and right part

%\item ``$\U(N)$"-simplicity constraints don't close anymore

%\item but holomorphic constraints still work. And still can be solved using the same ansatz! (e.g take $\det\Theta=...$ )

%\end{itemize}

The next step is to extend our construction to the Euclidean case with a finite Immirzi parameter $\gamma$ ($\gamma>0, \, \gamma\neq 1$). All the simplicity constraints (\ref{diag}) and (\ref{crosseddiag}) are then modified: at the discrete level, there is no equality between the left and the right parts of the scalar products anymore; the relation between the left and the right parts becomes a proportionality relation.  For any  two faces  $\Delta$, $ \tilde{\Delta}$  of a 3-cell:
\be
\vJ_\Delta^L\cdot \vJ_{\tilde{\Delta}}^L=\rho^2 \vJ_\Delta^R\cdot \vJ_{\tilde{\Delta}}^R
\ee
with $\rho >0$ and where the cases $\Delta=\tilde{\Delta}$ correspond to the diagonal simplicity constraints and the cases $\Delta \neq \tilde{\Delta}$ correspond to the cross-simplicity constraints. The proportionality coefficient $\rho$ is simply related to the Immirzi parameter $\gamma$ by: $\rho\equiv \f{\gamma+1}{|\gamma-1|}$ \cite{FK, EPRL}.
Once again, we would like to use the $\U(N)$ formalism to solve these constraints. We therefore focus on a $\Spin(4)$ intertwiner with $N$ legs labelled by $i\, \in \{1, \cdots, N \}$. The issue remains that the crossed simplicity constraints $\vJ_i^L \cdot \vJ_j^L- \rho^2 \vJ_i^R \cdot \vJ_j^R=0$ for all couples of legs $i \neq j$ do not form a closed algebra. Following the same idea as previously we would like to replace the simplicity constraints by simpler constraints expressed in term of the operators $E^{L,R}_{ij}$ or $F^{L,R}_{ij}$ and $(F_{ij}^{L,R})^\dagger$ which form a closed algebra. We tried all possible combinations of $E$ and $F$ constraints and the only way to get a closed algebra including all the simplicity constraints is to consider constraints of the form:
\bes
\cC_i&= &E_{i}^L-E_i^R =0, \quad \forall \, i, \quad \textrm{ for the diagonal simplicity constraints.} \nn \\
\cF_{ij}^{\rho}&\equiv& F_{ij}^L - \rho (F_{ij}^R)^{\dagger}=0 \quad \forall \, i, j \quad \textrm{ for the cross simplicity constraints}
\ees
Then,
\be
[ \cF_{ij}^\rho, \cF_{kl}^\rho]=0 \quad \textrm{ and } [\cC_i, \cF^\rho_{kl}]= \delta_{il}\cF^\rho_{ik}-\delta_{ik}\cF^\rho_{il}.
\ee
We can again define a Hilbert space $\cH^\rho$ of states satisfying these constraints: $\cC_i |\psi \ra=0 \; \forall i, \, \cF_{ij}^{\rho} |\psi \ra=0 \;  \forall i \neq j$.
We already have one solution to both $\cC_i$ and $\cF_{ij}^{\rho}$ constraints given by the state $|\alpha=\rho\ra$ as defined in the previous subsection by (\ref{alpha}).
However, we still have the usual un-rescaled diagonal simplicity constraint which do not involve the Immirzi parameter. Then, as for the crossed simplicity constraints, the result is also disappointing
and we have that $\forall |\psi \ra, \, |\phi \ra \in \cH^\rho$:
\bes
\la \psi | \vJ^L_i\cdot \vJ^L_j -\f14 E_i^LE_j^R | \phi \ra &= & \la \psi | -\f12 F_{ij}^{L \dagger} F_{ij}^L| \phi \ra \nn \\
&=& \rho^2 \la \psi | -\f12 F_{ij}^{R} F_{ij}^{R \dagger}| \phi \ra \nn \\
&=&\rho^2 \la \psi | \vJ^R_i\cdot \vJ^R_j -\f14 (E_i^R+2)(E_j^R+2) | \phi \ra.
\ees
Thus, since $E_i^L=E_i^R$, we get at the leading order in $j$ that the "right" observables $(\vJ_i^R\cdot \vJ_j^R -E_iE_j) \sim |J_i| |J_j|(\cos \theta_{ij}^R-1)$ are rescaled by the proportionality coefficient $\rho^2$ with respect to the "left" observables $\vJ_i^L\cdot \vJ_j^L -E_iE_j) \sim |J_i| |J_j|(\cos \theta_{ij}^L-1)$ where $\theta_{ij}$ is the angle between the two vectors $\vJ_i$ and $\vJ_j$. However, these observables which are corrected observables compared to the scalar product observables, do not have any real interesting geometrical interpretations and it does not seem possible to extract the expected relation: $\la \vJ_i^L \cdot \vJ_j^L \ra= \rho^2 \la \vJ_i^R \cdot \vJ_j^R \ra$.

Here again, it seems that the main obstacle is imposing strongly the diagonal simplicity constraints. In the following section, we will show how to relax the diagonal simplicity constraints and solve weakly all the simplicity constraints using coherent states for an arbitrary value of the Immirzi parameter.

\section{Weakening the Constraints}
%%%%%%%%%%%%%%%%%
In the previous section, we focused on the issue of the cross-diagonal simplicity constraints $\vJ_i^L\cdot \vJ_j^L-\vJ^R_i\cdot \vJ_j^R=0$ which have to be imposed weakly since they do not form a closed algebra. We defined some new sets of constraints $\{ \cC_{ij}, \, i<j \}$ or $\{ \cF_{ij} \}$ which allow to solve the cross simplicity constraints weakly and which are compatible with the diagonal simplicity constraints in such a way that  the sets of all constraints form a closed algebra and therefore can all be imposed strongly in a consistent way. This means that until now we tried to solve the crossed simplicity constraints weakly whereas the diagonal simplicity constraints were imposed strongly. Here, we propose to relax all simplicity constraints because there are in fact physically on an equal footing and there is no physical reason to deal with the diagonal simplicity constraints in a different way from the cross simplicity ones. The idea is to use coherent states to solve weakly all simplicity constraints in the semi-classical regime. We first go back to the usual $\SU(2)$ coherent states, then we will propose the $\U(N)$ coherent states that solve weakly all simplicity conditions for arbitrary Immirzi parameter.

%%%%%%%%%%
\subsection{Back to $\SU(2)$ coherent intertwiners}
%%%%%%%%%%%

The $\SU(2)$ coherent intertwiners $||j_i, \hat{n}_i \ra_L\otimes ||j_i, \hat{n}_i \ra_R$ are currently used to solve the simplicity constraints.  The usual analysis has been recalled in section \ref{SU2coh}. It is interesting to notice that these intertwiners are strong solutions to the diagonal simplicity constraints and that  there does not seem to exist any  other exact equation strongly solved by these states  in order to weakly solve the cross diagonal simplicity constraints even in the semi-classical regime.

In fact, $||j_i, \hat{n}_i \ra\otimes ||j_i, \hat{n}_i \ra= \int_{\Spin4} dG \, G\, \triangleright \otimes_{i=1}^N |2j_i, \hat{n}_i \ra$ span a Hilbert space of intertwiners which is the Hilbert space of intertwiners symmetric under the exchange of the left and right part. We denote it $\cH_{\textrm{sym}}$.
This symmetric Hilbert space $\cH_{\textrm{sym}}$ is generated by applying the operators
$E_{ij}^L E_{ij}^R$, $F_{ij}^LF_{ij}^R$, $F^{L \dagger}_{ij}F_{ij}^{R \dagger}$ on the vacuum state $|0\ra$.
It is obvious that any state $\psi \in \cH_{\textrm{sym}}$ satisfies all the non-diagonal simplicity constraints $\la \vJ_i^L\cdot \vJ_j^L\ra=\la \vJ^R_i\cdot \vJ_j^R\ra$ in expectation values. $\cH_{\textrm{sym}}$ is even the largest Hilbert space such that all the matrix elements of the constraints vanish:
\be
\forall \, \psi, \, \phi \in \cH_{\textrm{sym}}, \quad \la \psi |\vJ_i^L\cdot \vJ_j^L-\vJ^R_i\cdot \vJ_j^R| \phi \ra=0.
\ee

However, this symmetry property of the states does not seem to be fundamental. Indeed, we have seen in the section \ref{SU2coh} that there is a second sector solution to the cross simplicity constraints given by the states $||j_i, -\hat{n}_i \ra_L\otimes ||j_i, \hat{n}_i \ra_R$. These states are not symmetric anymore in the exchange of the left and right part but they clearly satisfy the simplicity constraints in expectation value. Moreover, the previous analysis is not generalizable to the case of Euclidean gravity with a finite Immirzi parameter $\gamma$: the cross simplicity constraints become  $\vJ_i^L\cdot \vJ_{j}^L=\rho^2 \vJ_i^R\cdot \vJ_{j}^R$ and thus, the symmetric intertwiners cannot be used  to solved them weakly anymore. The resolution done in \ref{SU2coh} is not generalizable when the Immirzi parameter is taken into account; usually the diagonal simplicity constraints are imposed strongly and the quadratic cross simplicity constraints are replaced by linear constraints $\vJ_i^L=\pm \rho \vJ_i^R$ which are then used to construct a so-called Master constraint in order to solve weakly the off-diagonal simplicity constraints \cite{EPRL}.

We will now see that it is in fact possible to keep the standard quadratic simplicity constraints and to solve weakly all the simplicity constraints for any finite value of the Immirzi parameter.

%%%%%%%%%%%
\subsection{The final proposal: using $\U(N)$ coherent states}
%%%%%%%%%%%%%

Following the coherent state approach to solving the simplicity constraints, we propose to use the $\U(N)$ coherent states instead of the usual $\SU(2)$ coherent intertwiners. As we have already reviewed earlier, a $\U(N)$ coherent state $|J, \{z_k\}\ra$ is labeled by the total area $J$ and the $N$ spinors $z_k$ which define the semi-classical geometry underlying the intertwiner state. Now, considering $\Spin(4)$-intertwiners, we consider tensor products of $\U(N)$ coherent states for both the left and right sectors, that is $|J_L, \{z^L_k\}\ra\otimes |J_R, \{z^R_k\} \ra$. We would like to relax all simplicity constraints. Since we also relax the diagonal simplicity constraints, we do not require the matching of the total areas of the left and right sectors and we work with a priori two different $\U(N)$ representations, $J_L\neq J_R$. Then, the simplicity constraints impose that the classical geometry of the left and right intertwiners  are the same up to an overall scale. This will translate into relations between the spinors of the left and right sectors, $z^L_k$ and $z^R_k$.

Let us start by recalling the norm of the $\U(N)$ coherent states and the expectation values (normalized by the norm) of the geometric observables on them:
\bes
&&\la J, \{z_k\}\,|\,J, \{z_k\}\ra = A(z)^{2J},\qquad
A(z)\,=\,\f12\sum_k \la z_k|z_k\ra \\
&&\la E_{ij}\ra
=
J\,\f{\la z_i|z_j\ra}{A(z)},\qquad\forall i,j \nn\\
&&\la \vJ_i\cdot\vJ_j\ra
=
\f14 \f{J^2}{A(z)^2}\,\vV(z_i)\cdot\vV(z_j)+
\f{J}{8\,A(z)^2} \left(\vV(z_i)\cdot\vV(z_j)-3|\vV(z_i)|\,|\vV(z_j)|\right),\qquad \forall i\ne j, \nn
\ees
%\bes
%\la J, \{z_k\}\,|\,J, \{z_k\}\ra &=& A(z)^{2J},\qquad
%A(z)\,=\,\f12\sum_k \la z_k|z_k\ra \nn\\
%\la E_{ij}\ra
%&=&
%J\,\f{\la z_i|z_j\ra}{A(z)},\qquad\forall i,j \nn\\
%\la \vJ_i\cdot\vJ_j\ra
%&=&
%\f14 J^2\,\vV(z_i)\cdot\vV(z_j)+
%\f J8 \left(\vV(z_i)\cdot\vV(z_j)-3|\vV(z_i)|\,|\vV(z_j)|\right),\qquad \forall i\ne j. \nn
%\ees
%
where we have implicitly assumed that the spinors $z_k$ satisfy the closure conditions. In case they do not close, the formulas above still hold up to replacing $A(z)$ by the the determinant $\sqrt{\det\, X(z)}$ as explained in the previous sections.

From these expressions, two things are clear. First, the total area label $J$ is simply a scale factor, it does not affect further the details of the classical geometry determined by the spinor labels. Thus, it appears that the ratio of the total area of the left and right sectors defines directly the Immirzi parameter $\rho=\f{J_L}{J_R}$. Second, if we want to match up to an overall factor the expectation values of the scalar product $\la \vJ_i\cdot\vJ_j\ra$ of the left and right sectors, it is clear that we have to require that the 3-vectors $\vec{V}(z_k)$ are the same up to a sign for the left and right sectors. Thus we distinguish two classes of solutions, which correspond to the two regimes, standard (s) and dual ($\star$), of simplicity constraints:
\begin{enumerate}

\item We require ${z^L_k= z^R_k}$ and consider the tensor product $|J_L, \{z_k\}\ra\otimes |J_R, \{z_k\} \ra$. This means that $\vec{V}(z_k^L)=\vec{V}(z_k^R)$. This corresponds to the {\bf standard simplicity regime (s)}. At leading order in the total area $J_{L,R}$, we have the equality of the expectation values of the scalar product observables:
    \be
    \la \vJ_i^L\cdot\vJ_j^L\ra \,\sim\, \rho^2\,\la \vJ_i^R\cdot\vJ_j^R\ra,
    \qquad \rho=\f{J_L}{J_R}.
    \ee
    Moreover, we also have the exact  equality of the expectation values of the $\u(N)$ generators:
    \be
    \la E_{ij}^L\ra =\rho  \la E_{ij}^R\ra.
    \ee
    There is still a $\U(N)$ action on the set of coherent states $|J_L, \{z_k\}\ra\otimes |J_R, \{z_k\} \ra$. Indeed the diagonal action $(u,u)$ acts simultaneously on the two sets of spinors, $(z_k,z_k)\arr ((u\,z)_k,(u\,z)_k)$. Thus these are still coherent states {\it \`a la} Perelomov.

\item We require {\bf $z^L_k= \varsigma z^R_k$} and consider the tensor product $|J_L, \{z_k\}\ra\otimes |J_R, \{\varsigma z_k\} \ra$. This means that $\vec{V}(z_k^L)=-\vec{V}(z_k^R)$ and corresponds to the {\bf dual simplicity regime ($\star$)}. At leading order in the total area $J_{L,R}$, we still have the equality of the expectation values of the scalar product observables:
    \be
    \la \vJ_i^L\cdot\vJ_j^L\ra \,\sim\, \rho^2\,\la \vJ_i^R\cdot\vJ_j^R\ra,
    \qquad \rho=\f{J_L}{J_R}.
    \ee
    However the equality of the expectation values of the $\u(N)$ generators is slightly modified due to the fact that the $\varsigma$ map is anti-unitary. Indeed, taking into account that $\la \varsigma z_i|\varsigma z_j \ra=\la z_j|z_i \ra=\overline{\la z_j|z_i \ra}$, we now have:
    \be
    \la E_{ij}^L\ra =\rho  \la E_{ji}^R\ra =\rho  \overline{\la E_{ij}^R\ra}\,.
    \ee
    The  $\U(N)$ action which is consistent with this set of coherent states $|J_L, \{z_k\}\ra\otimes |J_R, \{\varsigma z_k\} \ra$ is  the diagonal action $(u,\bar{u})$ which is actually generated by our $\u(N)$ simplicity condition $\cC_{ij}$ and which  acts simultaneously on the two sets of spinors as
    %$(z_k,\varsigma z_k)\arr ((u\,z)_k,u\,\varsigma z_k)=((u\,z)_k,\varsigma(u\,z)_k)$.
    $(u,\bar{u})\vartriangleright(z_k,\varsigma z_k)\,=\, ((u\,z)_k,\varsigma(u\,z)_k)$.
    Thus these are also coherent states {\it \`a la} Perelomov.

\end{enumerate}

Therefore, just like when using coherent intertwiners to solve weakly the simplicity constraints, we can clearly implement the two regimes of simplicity for the intertwiners. However, there are clear advantages of this new approach over the usual one. First, there are no big difference in the properties of the $\U(N)$ coherent states corresponding to the two sectors. Second, the $E_{ij}$ observables allow to easily distinguish the two sectors. Third, we have $\U(N)$ actions in both cases which allow consistently deform these intertwiners, thus endowing them with a true structure of coherent states and not mere semi-classical states.

\medskip

For the moment, we have managed to solve weakly both diagonal and cross simplicity constraints using the coherent states $|J_L, \{z_k^L\}\ra\otimes |J_R, \{z_k^R\} \ra$ with $z_k^L=z_k^R$ or $z_k^L=\varsigma z_k^R$. This provides solutions to the simplicity constraints for values of the Immirzi parameter corresponding to the ratio $\rho= J_L/J_R$. This parameter still takes discrete values. However, since we have decided to relax the diagonal simplicity constraints and thus {\it not} require an exact match between the individual spins $j_i^{L,R}$ of the left and right sectors, we can further relax our implicit assumption that the total area need to be fixed. Then we would only require a matching of the total areas of the left and right sectors in expectation value and the parameter $\rho= \la J_L\ra/\la J_R\ra$ will be allowed to take any (positive) real value.

To implement this, we come back to the $F$-constraints considered earlier in section \ref{Fconstraints} and in section \ref{immparam?}~:
\be
F_{ij}^L= \rho F_{ij}^R.
\ee
These constraints were not compatible with the diagonal simplicity constraints. However, since we have decided to relax these diagonal simplicity constraints, we can neglect them and impose the $F$-constraints strongly.
We can easily solve these constraints since we know how to diagonalize the annihilation operators $F_{ij}$. Indeed, a generic solution will be given by the tensor product of $\beta$-states:
\be
|\beta_L, \{z_k^L\}\ra\otimes |\beta_R, \{z_k^R\} \ra,
\qquad
\textrm{with}\quad
\beta_L=\rho\beta_R
\quad\textrm{and}\quad
z_k^L=z_k^R.
\ee
We remind the definition of the $\beta$-states as superpositions of coherent states for different values of the total area:
\be
|\beta,\{z_k\}\ra
\,=\,
\sum_J \f{\beta^{2J}}{\sqrt{J!(J+1)!}}\,|J,\{z_k\}\ra,
\ee
which satisfy the eigenvalue equation:
$$
F_{ij}\,|\beta,\{z_k\}\ra\,=\,
\beta^2Z_{ij}
\,|\beta,\{z_k\}\ra,\qquad
\textrm{with}\quad
Z_{ij}=(z^0_iz^1_j-z^1_iz^0_j).
$$
Once again, we can easily compute the norm of these states, as well as the expectation values of the geometric observables~:
\bes
&&\la \beta, \{z_k\}\,|\,\beta, \{z_k\}\ra = \f{I_1(2x)}{x},\qquad
\textrm{with}\quad x\,=\,|\beta|^2A(z), \\
&&\la E_{ij}\ra
=
\f{xI_2(2x)}{I_1(2x)}\,\f{\la z_i|z_j\ra}{A(z)},\qquad\forall i,j \nn\\
&&\la \vJ_i\cdot\vJ_j\ra
=
\f14 \f{\vV(z_i)\cdot\vV(z_j)}{A(z)^2}\,\f{x\left(\f32 I_2(2x)+xI_3(2x)\right)}{I_1(2x)}
-\f38\f{|\vV(z_i)|\,|\vV(z_j)|}{A(z)^2}\,\f{xI_2(2x)}{I_1(2x)},\qquad \forall i\ne j, \nn
\ees
where the $I_n$'s are the modified Bessel functions of the first kind and the parameter $x=|\beta|^2A(z)$ depends very simply on the label $\beta$.
For large values of $x$, i.e for large area $A(z)$ or large value of $\beta$ (this is more or less the same since the label $\beta$ can be entirely absorbed as a overall rescaling of the spinors $z_k$ in the definition of the $\beta$-states), these expressions simplify at leading order and we get~:
\bes
&&\la E_{ij}\ra
\sim
x\,\f{\la z_i|z_j\ra}{A(z)}
\sim
|\beta|^2\,{\la z_i|z_j\ra},
\qquad\forall i,j \\
&&\la \vJ_i\cdot\vJ_j\ra
\sim
\f{x^2} 4 \f{\vV(z_i)\cdot\vV(z_j)}{A(z)^2}
\sim
|\beta|^4  \f{\vV(z_i)\cdot\vV(z_j)}{4},
\qquad \forall i\ne j. \nn
\ees
Thus, considering tensor product states $|\rho \beta, \{z_k\}\ra\otimes |\beta, \{z_k\} \ra$ with $\beta_L=\rho\beta_R$ and $z_k^L=z_k^R$, we obtain exact solutions to the $F$-constraints $(F_{ij}^L-\rho F_{ij}^R)\,|\psi\ra=0$. And these solutions satisfy weakly the simplicity conditions at leading order in the semi-classical limit, $\la \vJ_i^L\cdot\vJ_j^L\ra\sim\rho^2\la \vJ_i^R\cdot\vJ_j^R\ra$ and $\la E_{ij}^L\ra\sim\rho\la E_{ij}^R\ra$.

We proceed similarly with the other sectors and consider tensor product states $|\rho \beta, \{z_k\}\ra\otimes |\beta, \{\varsigma z_k\} \ra$, with $\beta_L=\rho\beta_R$ and $z_k^R=\varsigma z_k^L$.  These solutions satisfy weakly the simplicity conditions at leading order in the semi-classical limit. Indeed, we have obviously $\la \vJ_i^L\cdot\vJ_j^L\ra\sim\rho^2\la \vJ_i^R\cdot\vJ_j^R\ra$, but $\la E_{ij}^L\ra\sim\rho\,\overline{\la E_{ij}\ra}$. However, the main difference is that we have not been able to identify a set of constraints as the $F$-constraints which would characterize these tensor states. Indeed, looking at the action of the $F_{ij}$ operators, we get:
\bes
&&F_{ij}^L\,|\rho \beta, \{z_k\}\ra\otimes |\beta, \{\varsigma z_k\} \ra
\,=\,
\rho\beta\,Z_{ij}\,|\rho \beta, \{z_k\}\ra\otimes |\beta, \{\varsigma z_k\} \ra,\nn\\
&&F_{ij}^R\,|\rho \beta, \{z_k\}\ra\otimes |\beta, \{\varsigma z_k\} \ra
\,=\,
\beta\,\overline{Z_{ij}}\,|\rho \beta, \{z_k\}\ra\otimes |\beta, \{\varsigma z_k\} \ra,\nn
\ees
and we actually don't know any operator which would act anti-holomorphically on states $|\beta, \{z_k\}\ra$ so as to produce the value $\overline{Z_{ij}}$. This is very similar to what happens when solving the simplicity constraints using the standard coherent intertwiners: the coherent intertwiner span a subspace in the standard regime (s) while they still span the whole Hilbert space of intertwiners in the dual regime ($\star$). However, our approach still has two very interesting advantages: the $\U(N)$ action on our solution states and the straightforward inclusion of the Immirzi parameter in our framework as a simple scale factor.

\medskip

We would like to finish this last section with a remark on the phase of the spinors. Indeed, the matching of the expectation values of the scalar product observables of the left and right sectors only requires a matching of the 3-vectors $\vec{V}(z^L_k)=\pm\vec{V}(z^R_k)$ with the sign depending on whether we are in the standard regime or the dual regime. In order to impose these equalities, we have required that $z^R_k=z^L_k$ or that $z^R_k=\varsigma z^L_k$. However, the 3-vector $\vec{V}(z)$ only determines the spinor $z$ up to a global phase, $z\arr e^{i\theta}\,z$. We can thus multiply any of the $2N$ spinors  $z^L_k$ and $z^R_k$ by arbitrary phases without affecting the expectation values $\la \vJ^L_i\cdot\vJ^L_j \ra$ and $\la \vJ^R_i\cdot\vJ^R_j \ra$. Therefore, we can consider generally coherent states $|\rho J, \{e^{i\theta_k^L}z_k\}\ra\otimes |J, \{e^{i\theta_k^R}z_k\} \ra$ with arbitrary phases  $\theta_k^L,\theta_k^R$. These tensor products will still solve weakly the quadratic simplicity constraints on the scalar product operators.
The expectation values of the $\u(N)$ generators $\la E_{ij}^{L,R}\ra$ are nevertheless sensitive to these phases and are equal only up a phase. Since the geometry of the 3-vectors $\vV(z_k^{L,R})$, and thus the geometry of the intertwiner, do not depend on the phases of the spinors, it is natural to wonder about their physical/mathematical relevance.

The answer proposed in \cite{UN3} is that these phases are relevant to the spin network construction when we glue intertwiners together. Indeed, following the interpretation of loop quantum gravity in term of discrete twisted geometries \cite{twisted}, these phases (or more precisely the relative phase between two intertwiners glued along an edge) encode the extrinsic curvature at the discrete level. In our context, having these freedom in shifting these phase without affecting the intrinsic geometry of the intertwiner (defined in term of the 3-vectors) should allow to glue these $\U(N)$ coherent intertwiners in a consistent way without interfering with the simplicity constraints.

%%%%%%%%%
\section{Conclusion and Outlook}
%%%%%%%%%

In the spinfoam approach, the simplicity constraints, which turn the SO(4) BF theory into 4d Euclidean gravity theory, are discretized and have to be imposed on the $\Spin(4)$-intertwiners from which are built the quantum states of geometry and the spinfoam transition amplitudes. The issue to implement the simplicity constraints without freezing too many local degrees of freedom comes from the fact that they do not form a closed algebra at the discrete level and cannot be imposed strongly.

The purpose of this paper has been to revisit the implementation of the discrete simplicity constraints using the $\U(N)$ framework initially developed for $\SU(2)$-intertwiners  in \cite{UN1, UN2, UN3}. Based on the Schwinger representation of the $\su(2)$ Lie algebra in term of a couple of harmonic oscilaltors, this framework introduces a new set of $\SU(2)$-invariant operators acting on the space of $\SU(2)$-intertwiners. These operators act on pairs of legs $(i,j)$ of the intertwiners: $E_{ij}$ generates $\U(N)$ transformations that deform the shape of the intertwiner, while $F_{ij}$ and $F\dag_{ij}$ act as annihilation and creation operators consistent with the $\U(N)$-action. The key result of this approach is that these $\SU(2)$-invariant observables form a closed algebra.
In the spinfoam context, we deal with the $\Spin(4)$-intertwiners. Using the decomposition of $\Spin(4)=\SU(2)_L\times\SU(2)_R$ in left and right sectors, we now have invariant operators acting on both sectors $E_{ij}^{L,R},\, F_{ij}^{L,R}, \, F_{ij}^{L,R}{}\dag$ which can be used to investigate how to impose the simplicity constraints.
More precisely, the idea developed in this paper is to recast the discrete simplicity constraints in term of observables defined in term of the $\U(N)$ operators and that form closed algebra. At the end of the day, it allows us to propose a set of $\U(N)$ coherent states that solve the simplicity constraints weakly at large scale for arbitrary values of the Immirzi parameter.

In the first part of this paper, we have completed the analysis of the $\U(N)$ framework for $\SU(2)$-intertwiners initiated in \cite{UN1, UN2, UN3}. We reviewed the $\U(N)$ coherent states introduced in \cite{UN3}. For a $N$-valent $\SU(2)$-intertwiner, they are labeled by  the total area $J=\sum_i j_i$ and  a set of $N$ spinors $z_k$. These coherent states $|J, \{z_k\}\ra$ form a over-complete basis for the space of $\SU(2)$-intertwiners at fixed area $J$ and are simply related to the Livine-Speziale coherent intertwiners. Moreover, we give explicitly the action of the SU(2) invariant operators on these $\U(N)$ coherent states as differential operators.

%These SU(2) invariant operators which act  on the space of SU(2) intertwiners are the $\U(N)$ generators $E_{ij}$ which allow to deform the shape of the intertwiners at fixed area $J$, and the $F_{ij}$, $F^\dagger_{ij}$ operators which allow to modify the total area $J$ -- the annihilation $F_{ij}$ decrease the total area $J$ by $-1$ whereas the creation operators  $F_{ij}^\dagger$ increase it by $+1$. One of the interesting property of these operators is that they form all together closed algebra.

In the second part of this paper, we have applied these new $\U(N)$ tools to the analysis of the simplicity constraints for $\Spin(4)$-intertwiners. The simplicity constraints couple the left and right sectors of the intertwiners. We have focused in re-expressing them in term of the $E,F,F\dag$ operators of the $\U(N)$ formalism. Following the usual approach, we have always distinguished the diagonal simplicity constraints from the cross simplicity constraints. The diagonal constraints act on single legs of the intertwiner and require that the $\Spin(4)$-representation living on a leg $i$ be simple i.e that the left and right spins are equal $j_i^L=j_i^R$ (or $j_i^L=\rho j_i^R$ for a non-trivial Immirzi parameter). These diagonal constraints are always imposed strongly on the intertwiner states. On the other hand, the cross simplicity constraints deal with pairs of legs and are standardly solved weakly in the most recent spinfoam models i.e only in expectation value (with minimal uncertainty).
We started by showing that the discrete simplicity constraints which do not form a closed algebra can be replaced by new constraints $\cC_{ij}$ which form a $\u(N)$-algebra. These new $\u(N)$ simplicity constraints are very simply constructed in term of the $E$-operators. We also explored other possibilities of constraint operators based on the operators $F$ and $F\dag$. In the end, it appeared that distinguishing the diagonal constraints from the cross constraints and imposing the first strongly while solving the later only weakly always lead to difficulties.

Thus, in the last part of our work, we propose to put all (diagonal and cross) simplicity constraints on the same footing and solve all of them at once in a weak way. This lead us to introduce constraints $F_{ij}^L-F_{ij}^R=0$ involving only annihilation operators. These constraints can be considered as the holomorphic constraints of the Gupta-Bleuler quantization procedure. Solving them in term of $\U(N)$ coherent states provides us with weak solutions to all simplicity constraints, for arbitrary values of the Immirzi parameter.

The next important question to explore is how to generalize this framework to  the Lorentzian case in order to check whether it is also possible to construct coherent states which could solve all simplicity constraints with an arbitrary Immirzi parameter. Another issue is to understand how to glue these $\U(N)$ coherent intertwiners consistently into spin network states in order to generalize our analysis to triangulations formed of an arbitrary number of polyhedra glued together. Finally, we hope that the introduction of these $\U(N)$ coherent states as a basis of the boundary physical Hilbert space of spinfoam model could help to understand the symmetries of the spinfoam amplitudes and their behavior under (discrete) deformations or diffeomorphisms.

%%%%%%%%%
%\section*{Acknowledgements}
%%%%%%%%%

%EL is partially supported by the ANR ``Programme Blanc" grants LQG-09.

%%%%%%%%%%%%%%%%%%%%%%%%%%%%%%%%%%%%%%%%%%%%%%
\appendix

%%%%%%%%%
\section{Coherent States for the Harmonic Oscillator} \label{cohHO}
%%%%%%%%%

Let us review the standard definition of coherent states for a single harmonic oscillator, defined by its creation and annihilation operators satisfying the commutation relation $[a,a\dag]=1$. The standard basis is defined by the number of quanta:
\be
a|n\ra=\,\sqrt{n}\,|n-1\ra,\qquad
a\dag|n\ra=\,\sqrt{n+1}\,|n+1\ra,\qquad
a\dag a|n\ra\,=\,n|n\ra.
\ee
Coherent states are defined through a sum over the standard basis:
\be
|z\ra= \sum_n \f{z^n}{\sqrt{n!}}\,|n\ra
= \sum_n \f{z^n}{\sqrt{n!}}\,\f{(a\dag)^n}{\sqrt{n!}}\,|0\ra
= e^{z\,a\dag}\,|0\ra.
\ee
This definition is not normalized, but we can easily compute its norm and define normalized states:
\be
\la z|z\ra= e^{|z|^2},\qquad
|z\ra_N\,\equiv\, e^{-\f{|z|^2}{2}}\,|z\ra.
\ee
The action of the $a,a\dag$ operators can be derived directly from the definition of the coherent states as series. The coherent states diagonalize the annihilation operator $a$ while the creation operator $a\dag$ acts as a derivation:
\be
a\,|z\ra=z\,|z\ra,\quad
a\dag\,|z\ra
%=\sum_n \f{z^n}{\sqrt{n!}}\,\sqrt{n+1}\,|n+1\ra
=\sum_{n\ge1} n\f{z^{n-1}}{\sqrt{n!}}\,|n\ra
=\pp_z\,|z\ra.
\ee
This action can be straightforwardly on the normalized coherent states. Then we get a anti-holomorphic shift in the $a\dag$ action:
\be
a\,|z\ra_N=z\,|z\ra_N,\quad
a\dag\,|z\ra_N
%=\pp_z\,\left(e^{-\f{|z|^2}{2}}\,|z\ra\right)
=\pp_z\,e^{-\f{|z|^2}{2}}\,|z\ra
%=\left(e^{-\f{|z|^2}{2}}\pp_z-\f{\bar{z}}{2}e^{-\f{|z|^2}{2}}\right)\,|z\ra
=\left(\pp_z-\f{\bar{z}}{2}\right)\,|z\ra_N.
\ee
The coherent states naturally provides an over-complete basis and a new decomposition of the identity:
\bes
\int \f{d^2z}{\pi}\, |z\ra_N{}_N\la z|
&=&
\int \f{d^2z}{\pi}\, e^{-|z|^2}|z\ra\la z|
\,=\,
\sum_{m,n}\f{|m\ra\la n|}{\sqrt{m!}\sqrt{n!}}\,\int \f{d^2z}{\pi}\, e^{-|z|^2}\bar{z}^nz^m \nn\\
&=&
\sum_{m,n}\f{|m\ra\la n|}{\sqrt{m!}\sqrt{n!}}
\int_0^{+\infty} dr\,e^{-r^2}r^{m+n+1} \int_0^{2\pi} \f{d\theta}{\pi}\, e^{i(m-n)\theta}\nn\\
&=&
\sum_{n}\f{2}{n!}|n\ra\la n|\,
\int_0^{+\infty} dr\,e^{-r^2}r^{2n+1}
\,=\, \id.
\ees
%
%It is also possible to prove this equality by an alternative approach, by checking that the operator defined by the integral over $z$ of the projection over the coherent states commutes with $a\dag$~:
%%\be
%%a\,\int \f{d^2z}{2\pi}\, e^{-|z|^2}|z\ra\la z|
%%\,=\,
%%\int \f{d^2z}{2\pi}\, ze^{-|z|^2}|z\ra\la z|
%%\,=\,
%%\int \f{d^2z}{2\pi}\, -\pp_{\bar{z}}e^{-|z|^2}|z\ra\la z|\,
%%\,=\,
%%\int \f{d^2z}{2\pi}\, e^{-|z|^2}|z\ra\la z|\,a.
%%\ee
%\be
%a\dag\,\int \f{d^2z}{2\pi}\, e^{-|z|^2}|z\ra\la z|
%\,=\,
%\int \f{d^2z}{2\pi}\, -(\pp_ze^{-\bar{z}z})|z\ra\la z|
%\,=\,
%\int \f{d^2z}{2\pi}\, \bar{z}e^{-|z|^2}|z\ra\la z|\,
%\,=\,
%\int \f{d^2z}{2\pi}\, e^{-|z|^2}|z\ra\la z|\,a\dag.
%\ee
%Since that operator is Hermitian, it also commutes with the annihilation operator $a$. Since it commutes with $a$ and $a\dag$, it must necessarily be proportional to the identity. Finally we commutes its square:
%\be
%\int \f{d^2z}{2\pi}\, e^{-|z|^2}|z\ra\la z|\dots
%\ee
%
We can also check explicitly that the action of $a\dag$ on coherent states is correctly given by the adjoint of the action of the annihilation operator $a$:
\bes
\int [d^2z d^2w]\,\overline{\phi(z)}\psi(w) \la z|a\dag w\ra
&=&
-\int [d^2z d^2w]\,\overline{\phi(z)}\pp_w\psi(w) \la z| w\ra
\,=\,
\int [d^2z d^2w]\,\overline{\phi(z)}\psi(w)\pp_w\left(e^{\bar{z}w}\right)\nn\\
&=&
\int [d^2z d^2w]\,\bar{z}\overline{\phi(z)}\psi(w) \la z| w\ra
\,=\,
\int [d^2z d^2w]\,\overline{\phi(z)}\psi(w) \la az| w\ra.
\ees
Finally, these coherent states transform consistently under the $\U(1)$-action generated by the number of quanta operator $a\dag a$~:
\be
e^{i\tau a\dag a}\,|z\ra
\,=\,
\sum_n \f{z^n}{\sqrt{n!}}\,e^{i\tau n}\,|n\ra
\,=\,
|e^{i\tau}z\ra.
\ee

%%%%%%%%%
\section{Commutation Relations Of the $E,F,F\dag$ Action on Coherent States}
%%%%%%%%%

The commutation relations between these $F$, $F^\dagger$ and $E$ operators acting on the U(N) coherent states are straightforward to check:
\bes
\left[ E_{ij}, E_{kl}\right] |J, \{z_q\}\ra
&= & \delta^z_{kl}\left( \delta^z_{ij}\left( |J,\{z_k\} \ra \right) \right) - \delta^z_{ij}\left( \delta^z_{kl}\left( |J,\{z_k\} \ra \right) \right)\nn\\
&=&\left( \delta_{kj} \delta^z_{il}- \delta_{il} \delta^z_{kj} \right) |J,\{z_q\} \ra= \left( \delta_{jk}E_{il}- \delta_{il} E_{kj} \right) |J,\{z_q\}\ra ,
\ees
\bes
\left[E_{ij}, F_{kl} \right] |J, \{z_q\} \ra
&=&\sqrt{J(J+1)}\left( Z_{kl} \delta^z_{ij} \left( |J-1,\{z_q\} \ra \right) - \delta^z_{ij}\left( Z_{kl} |J-1, \{z_q\}\ra \right)\right)\nn\\
&=&\sqrt{J(J+1)} \left(\delta_{il} Z_{jk}- \delta_{ik}Z_{jl} \right) |J-1,\{z_q\}\ra \nn \\
&=& \left(\delta_{il} F_{jk}- \delta_{ik}F_{jl} \right) |J,\{z_q\}\ra ,
\ees
\bes
\left[E_{ij}, F_{kl}^\dagger \right] |J,\{z_q\}\ra
&= &\f{1}{\sqrt{(J+1)(J+2)}} \left( \Delta^z_{kl} \left( \delta_{ij}^z (|J+1,\{z_q\}\ra)\right)- \delta^z_{ij} \left( \Delta^z_{kl} (|J+1,\{z_q\} \ra ) \right) \right) \nn \\
&=&\f{1}{\sqrt{(J+1)(J+2)}}\left( \delta_{kj} \Delta^z_{il}-\delta_{lj} \Delta^z_{ik}\right) |J+1,\{z_q\}\ra \nn \\
&= &\left(\delta_{kj} F_{il}^\dagger- \delta_{lj}F_{ik}^\dagger \right) |J,\{z_q\}\ra ,
\ees
\bes
\left[ F_{ij}, F^\dagger_{kl}\right]|J,\{z_q\}\ra
&= &\Delta^z_{kl}\left(Z_{ij} |J,\{z_q\}\ra \right)- Z_{ij} \Delta_{kl}^z\left( |J,\{z_q\} \ra \right)\nn\\
&=& \left(\delta_{ki} \delta^z_{lj}-\delta_{kj} \delta^z_{li} -\delta_{li} \delta^z_{kj}+ \delta_{lj} \delta^z_{ki} + \Delta^z_{kl}(Z_{ij}) \right)|J,\{z_q\}\ra \nn \\
&=& \left(\delta_{ki} E_{lj}-\delta_{kj} E_{li} -\delta_{li} E_{kj}+ \delta_{lj} E_{ki} + 2(\delta_{ki}\delta_{lj}-\delta_{li}\delta_{kj}) \right)|J,\{z_q\}\ra .
\ees

%%%%%%%%%%%%%%
\section{Norm of the $\U(N)$-invariant state: $|J\ra$} \label{norm_App}
%%%%%%%%

Following the previous work done in \cite{2vertex}, we compute the norm of the $\U(N)$-invariant state $|J \ra= (f^\dagger)^J |0\ra$ where we have introduced the operator:
\be
f^\dagger= \sum_{kl} F_{kl}^{L \dagger } F_{kl}^{R \dagger}.
\ee
The norm of $|J\ra$ is then given by:
\be
\la 0 | f^J( f^\dagger)^J |0\ra
\ee
with $f= \sum_{kl} F_{kl}^{L } F_{kl}^{R }$. We need to determine the action of $f$ on $|J\ra$: $f |J\ra= f (f^\dagger)^J |0\ra$.
In order to compute this action of $f$ on $|J\ra$, we calculate the commutator between $f$ and $f^\dagger$:
\be
[f, f^\dagger]= 4 e \left(\f{E^L+E^R}{2} \right) + 4 \left( \f12 (E^L+E^R)^2+(E^L+E^R)(2N-1)+2N(N-1) \right)
\ee
where we have defined:
\be
e= \sum_{kl} E_{kl}^L E_{kl}^R
\ee
and we recall that $E^{L/R}=\sum_i E_i^{L/R}$. In our case of interest $E^L=E^R=E$. Moreover, to compute this commutator we used the fact that on the intertwiner space ($\vJ^L=\vJ^R=0$), the $F$ and $F^\dagger$ operators satisfy additional quadratic constraints:
\be
\sum_k \left(F^{L/R}_{ki}\right)^ \dagger F_{kj}^{L/R} = E_{ij}^{L/R} \left( \f{E^{L/R}}{2}+1\right), \qquad \sum_k F^{L/R}_{kj} \left(F_{ki}^{L/R}\right)^\dagger = (E_{ij}^{L/R}+2 \delta_{ij}) \left( \f{E^{L/R}}{2}+N-1\right).
\ee
We also need to compute the commutator between $e$ and $f^\dagger$. We use the fact that the $E$ and $F^\dagger$ also satisfy quadratic constraints on the intertwiner space:
\be
\sum_k \left(F^{L/R}_{ik}\right)^\dagger E_{jk}^{L/R} =  \left(F^{L/R}_{ij}\right)^\dagger \f{E^{L/R}}{2}, \qquad \sum_k E_{jk}^{L/R} \left(F^{L/R}_{ik}\right)^\dagger =  \left(F^{L/R}_{ij}\right)^\dagger \left( \f{E^{L/R}}{2} +N-1 \right),
\ee
then,
\be
[e, f^\dagger]= 2 f^\dagger \left(\f{E^L+E^R}{2}+N-1 \right)
\ee
where once again, $E^L$ and $E^R$ can be replaced by $E$. Moreover, this total area operator is clearly diagonal in the basis $|J\ra$:
\be
E |J\ra = 2J |J\ra
\ee
We can then deduce the action of $f$ on $|J \ra$:
\bes
f(f^\dagger)^J|0\ra&=&\sum_{k=0}^{J-1}\left\{4(2(J-1-k)+N) (f^\dagger)^k e \, (f^\dagger)^{J-1-k} |0\ra + 16 (2 (J-1-k)^2 +(J-1-k)(2N-1)) (f^\dagger)^{J-1}|0\ra \right\} \nn\\
&& \qquad \qquad+8 (J-1)N(N-1) (f^\dagger)^{J-1}|0 \ra \nn \\
&=&[\sum_{k=0}^{J-1}\left\{ 8(2(J-1-k)+N)(J-1-k)(J+N-k-3)+ 16 (2 (J-1-k)^2 +(J-1-k)(2N-1))\right\}  \nn \\
&& \qquad \qquad +8 (J-1)N(N-1)] |J-1\ra \nn \\
&=& 4J(J+1)(N+J-1)(N+J-2) |J -1 \ra
\ees
Using this action, we compute the norm of the state $|J \ra$ by recursion:
\be
\la J | J\ra = \la 0 | f^{J-1} f |J\ra= 4J(J+1)(N+J-1)(N+J-2) \la J-1 | J-1 \ra
\ee
which leads us to the scalar product:
\be
\la J | J\ra= 2^{2J} J! (J+1)! \f{(N+J-1)!(N+J-2)!}{(N-1)!(N-2)!}
\ee
 %%%%%%%%%%%%%%%%%%%%%%%%%%%%%%%%%%%%%%%

\end{document}